\documentclass[10pt,aps,prl,fleqn,superscriptaddress,twocolumn,nofootinbib,preprintnumbers]{revtex4-2}
\pdfoutput=1
\usepackage{hyperref,amsmath,amssymb,nicematrix,graphicx,nicefrac}
\hypersetup{
  pdfauthor={Enrico Bothmann,Timo Janßen,Max Knobbe,Bernhard Schmitzer, Fabian Sinz},
  pdftitle={Monte Carlo Event Generation with Continuous Normalizing Flows}
}

\usepackage{tabularray}
\UseTblrLibrary{siunitx}
\usepackage{xfrac}

\usepackage{mathtools}
\usepackage{float}
\usepackage{placeins}

\usepackage{todonotes} 
\newcommand{\acronymsw}[1]{\textsc{#1}}
\newcommand{\Pepper}{\acronymsw{Pepper}}
\newcommand{\Chili}{\acronymsw{Chili}}

\newcommand{\Vegas}{\acronymsw{Vegas}}
\newcommand{\Sherpa}{\acronymsw{Sherpa}}

\newcommand{\Pythia}{\acronymsw{Pythia}}
\newcommand{\NNPDF}{\acronymsw{NNPDF}}
\newcommand{\LHAPDF}{\acronymsw{LHAPDF}}

\newcommand{\acronym}[1]{#1}
\newcommand{\LHC}{\acronym{LHC}}

\newcommand{\dif}{\mathop{}\!\mathrm{d}}

\makeatletter
\newcommand{\spx}[1]{%
  \if\relax\detokenize{#1}\relax
    \expandafter\@gobble
  \else
    \expandafter\@firstofone
  \fi
  {^{#1}}%
}
\makeatother

\newcommand\pd[3][]{\frac{\partial\spx{#1}#2}{\partial#3\spx{#1}}}

\newcommand{\symscr}[1]{\mathcal{#1}}

\DeclarePairedDelimiter\abs{\lvert}{\rvert}

\DeclarePairedDelimiter\intervalrightopen{[}{)}

\begin{document}
\preprint{FERMILAB-PUB-25-0468-T}
\title{Monte Carlo Event Generation with Continuous Normalizing Flows}

\author{E.\ Bothmann}
\affiliation{IT Department, CERN, 1211 Geneva 23, Switzerland}
\affiliation{Institute for Theoretical Physics, University of Göttingen, Germany}

\author{T.\ Janßen}
\affiliation{Institute for Theoretical Physics, University of Göttingen, Germany}
\affiliation{Campus Institute Data Science, University of Göttingen, Germany}

\author{M.\ Knobbe}
\affiliation{Fermi National Accelerator Laboratory, USA}

\author{B.\ Schmitzer}
\affiliation{Campus Institute Data Science, University of Göttingen, Germany}
\affiliation{Institute for Computer Science, University of Göttingen, Germany}

\author{F.\ Sinz}
\affiliation{Campus Institute Data Science, University of Göttingen, Germany}
\affiliation{Institute for Computer Science, University of Göttingen, Germany}

\begin{abstract}
  We apply Continuous Normalizing Flows trained with the Flow Matching method
  to the problem of
  phase-space sampling in Monte Carlo event generation for high-energy collider physics.
  Focusing on lepton-pair and top-quark pair production with multiple jets, the two
  computationally most expensive processes at the Large Hadron Collider, we train
  helicity-conditioned Continuous Normalizing Flows to remap the random numbers used in matrix element evaluation.
  Compared to standard methods, we achieve unweighting efficiency improvements by factors of up to 184 and 25
  for the two processes at their respective highest jet number,
  at the cost of an increased evaluation time.
  When combining the advantages of Continuous Normalizing Flows
  with the fast evaluation times of Coupling Layer based Flows,
  using the RegFlow approach,
  we find parton-level unweighted event generation walltime gains
  of about a factor of ten at the highest jet numbers.
  These substantial gains
  highlight the promise of samplers based on machine learning for next-generation collider experiments.
\end{abstract}

\maketitle
\textit{Introduction}---The exceptional experimental
precision achieved by the ATLAS~\cite{ATLAS:2008xda} and CMS~\cite{CMS:2008xjf} experiments during the current and upcoming runs at the Large Hadron Collider (LHC),
and its potential successors~\cite{FCC:2018byv,FCC:2018evy,FCC:2018vvp}
demand equally precise
theoretical simulations.
Without any improvements over current techniques, the uncertainties in many measurements will be dominated
not by experimental statistics, but by deficiencies in Monte Carlo (MC) event samples that underpin critical
analyses~\cite{Buckley:2011ms,EuropeanStrategyGroup:2020pow,Narain:2022qud,HEPSoftwareFoundation:2017ggl,HSFPhysicsEventGeneratorWG:2020gxw}.
An important example is vector boson plus jets production,
for which very large samples are an essential input for many precision measurements,
e.g.\ for Higgs boson~\cite{ATLAS:2019vrd,ATLAS:2020jwz}
and top quark measurements~\cite{ATLAS:2019zrq,ATLAS:2020aln}.
These simulations require improvements in three main areas:
parametric accuracy, numerical stability, and
statistical precision.
This work addresses the third,
i.e.~how to efficiently generate large-scale MC event samples with a statistical quality that matches or exceeds that of experimental data samples.
The ATLAS Collaboration estimates that approximately 330 billion simulated events will be needed in the next phase of the LHC to accurately model vector-boson production in association with additional jets~\cite{ATLAS:2021yza}, a dominant background process in high-energy analyses.
Generating this dataset using current methods would demand
about
1000 Perlmutter 2$\times$CPU nodes for an entire year~\cite{ATLAS:2021yza}.

The main bottleneck is evaluating matrix elements for hard-scattering processes, especially at high final-state multiplicities.
Although only computationally feasible at leading order, these processes dominate the cost due to their complexity and low unweighting efficiency $\epsilon$~\cite{Bothmann:2022thx}.
The term unweighting refers to using rejection sampling
in order to replace a weighted sample with a typically much smaller
unit-weight sample that follows the same underlying probability distribution.
This is used in high-energy physics to reduce the cost of storage
and downstream simulation steps.
The unweighting efficiency $\epsilon$
is given by the number of events in the unweighted sample
divided by the number of events in the original weighted sample.
The main event generators used by LHC experiments are based on multi-channel methods~\cite{Kleiss:1994qy} and adaptive importance sampling algorithms like \Vegas~\cite{Lepage:1977sw,Ohl:1998jn,Lepage:2020tgj}. For seven final-state particles, one typically finds $\epsilon < \SI{0.01}{\percent}$~\cite{Hoche:2019flt}.

Modern machine learning techniques offer an alternative.
In particular, Normalizing Flows (NFs)~\cite{Tabak:2010,Tabak:2013,Dinh:2014} have been studied as drop-in replacements for \Vegas, offering more flexible function approximations~\cite{Klimek:2018mza,Bothmann:2020ywa,Gao:2020zvv,Heimel:2022wyj,Verheyen:2022tov,Heimel:2023ngj,Heimel:2024wph,Butter:2022rso}.
Improvements of up to a factor of 10 in efficiency were demonstrated for low to moderate final-state multiplicities. However, for the highest-multiplicity processes that dominate computational budgets, substantial gains have remained elusive.

In this work, we propose---for the first time---the use
of Continuous Normalizing Flows (CNFs)~\cite{Chen:2018}
trained with the Flow Matching method~\cite{Lipman:2023,Albergo:2023building,Albergo:2023stochastic,Liu:2022}
to solve this problem.
We evaluate our approach on the two most computationally intensive processes simulated for the LHC: lepton-pair production
and top-quark pair production
with multiple jets.
We compare performance
using the unweighting efficiency $\epsilon$
as our key metric,
benchmarking against \Vegas-based methods and Normalizing Flows based on Coupling Layers~\cite{Dinh:2014,Mueller:2019,Durkan:2019}.
Our results demonstrate a significant increase in $\epsilon$: for the highest-multiplicity processes, CNFs improve $\epsilon$ by factors of up to 184, compared to the other methods.
This is enabled in part by conditioning on helicity configurations, which allows the model to learn correlations between discrete and continuous features.

\textit{Phase-space sampling}---
In collider simulations, the primary quantity of interest is the scattering cross section, which measures the probability of a given scattering process to occur.
For hadronic collisions, e.g.\ at the LHC, it is given by
\begin{equation}
  \label{eq:hadronic_cross-section1}
  \begin{split}
    \sigma_{h_1h_2\to X} &= \sum_{i,j} \int_0^1 \!\dif x_1 \int_0^1 \!\dif x_2 \  f_i(x_1,
    \mu_F)\,  f_j(x_2, \mu_F) \\
                          &\qquad\times\, \hat\sigma_{ij \to X}(x_1, x_2, \mu_R,\mu_F)\,,
  \end{split}
\end{equation}
where the sum runs over partons in the incoming hadrons \(h_{1,2}\),
i.e.\ over quarks, antiquarks, and gluons. 
The parton density functions (PDFs) \(f_{i,j}\) are usually provided as interpolation grids in \(x\) and \(\mu_F^2\)~\cite{Buckley:2014ana}.
They vary non-linearly over many orders of magnitude.
The partonic cross section \(\hat\sigma_{ij\to X}\) involves an integral over the final-state four-momenta $p_f=(E_f,\vec p_f)$
with $f=3,\ldots,m$ and $m$ being the total number of incoming and outgoing particles:
\begin{equation}
  \label{eq:qft_cross-section1}
  \begin{split}
  \mathrm{d}\hat\sigma_{ij\to X} 
  &= \frac{1}{2 E_1 E_2 \abs{\vec{v}_1-\vec{v}_2}} \, \Big( \prod_f
  \frac{\mathrm{d}^3 \vec{p}_f}{(2\pi)^3} \frac{1}{2E_f} \Big) \\
  &\quad\;\,\times \Big|\symscr{M}_{ij\to X}\big(p_1,p_2 \to \{p_f\}\big)\Big|^2 \\ 
  &\quad\;\,\times \, (2\pi)^4 \, \delta^{(4)}\Big(p_1+p_2-\sum_{k=3}^m p_k\Big) \Theta\left(\left\{p_f\right\}\right) \,.
  \end{split}
\end{equation}
The cut function \(\Theta\), implemented in terms of Heaviside functions,
enforces phase-space cuts imposed by experiment or theory. The squared
matrix element \(\abs{\symscr{M}_{ij\to X}}^2\)
can be expressed via helicity amplitudes.
The Lorentz-invariant phase space includes a delta function enforcing four-momentum conservation, which reduces the dimensionality of the integral from \(3n_\text{out}\) to \(3n_\text{out} - 4\), with \(n_\text{out} = m - 2\) denoting the number of final-state particles.
Including \(x_{1,2}\), the total dimensionality becomes \(d = 3n_\text{out} - 2\), which can reach around 20 for typical LHC simulations.

The integral \(\sigma\) in eq.~\eqref{eq:hadronic_cross-section1} defines the overall normalization and can be
approximated to the required precision by MC integration.
While the relative error of this integral estimate is also of interest,
the real challenge is to generate phase-space
samples \(x_{1,2} \cup \{\vec{p}_f\}\) distributed as \(\mathrm{d}\sigma\).
These samples must contain enough events for robust comparisons across
a large number of observables of interest,
ranging across many orders of magnitude in $\mathrm d \sigma$.
The difficulty of the task is increased
because the matrix elements \(|\symscr{M}|^2\) are usually sharply
peaked and multimodal,
and because of the discontinuities introduced by the phase-space cuts \(\Theta\).
Moreover, the integration variables are often strongly correlated. 

To simplify the notation, let \(p\) be a probability density function defined by \(p(x) \dif x = \sigma^{-1} \dif
\sigma\). 
To make the problem more suitable for MC sampling,
the function \(p\) is pulled back
from the manifold of physically valid phase-space points, $M$, to the unit hypercube \(U = \intervalrightopen{0,
1}^d\) using a bijective map \(\phi: U \to M\), such that
\begin{equation}
    p_0(x) = \phi^* p(x) = p\bigl(\phi(x)\bigr) \det\biggl[{\pd{\phi}{x}}(x)\biggr] \,.
\end{equation}
The map \(\phi\) implements four-momentum conservation and Lorentz invariance. Ideally, it simplifies the problem by
using physical knowledge of the scattering process under consideration, in particular by flattening the
distribution.
Rejection sampling is used to sample from \(p\).
First, a trial event is generated by sampling from the uniform
distribution \(u_d\) on \(U\) and applying the map \(\phi\). The density of such events is
\begin{equation}
  q(x) = \phi_* u_d(x) = \det\biggl[ \pd{\phi^{-1}}{x} (x) \biggr] \,.
\end{equation}
The trial event is accepted with probability
\begin{equation}\label{eq:is_weight}
  p_{\text{accept}}(x) = \frac{w(x)}{w_\text{max}}\,, \quad \text{where} \quad w(x) \coloneq \frac{p(x)}{q(x)} \,,
\end{equation}
and rejected otherwise. Using \(w_{\text{max}} \coloneq \max_x w(x)\), the accepted events follow \(p\). 
The resulting unweighting efficiency~$\epsilon$
is given by the average acceptance probability,
\begin{equation}\label{eq:unw_eff}
  \epsilon \coloneq \frac{\langle w\rangle}{w_{\text{max}}}\,,
\end{equation}
which is a crucial figure of merit to measure the performance of a phase-space generator. 

In practice, the maximal weight $w_{\mathrm{max}}$ needs to be estimated in an initial survey phase from a finite set of events.
To mitigate the impact of large outliers,
one usually defines an effective $w_{\mathrm{max,eff}}$ to increase the efficiency and to render its value more numerically stable. 
Events with a weight $w>w_{\mathrm{max,eff}}$ are assigned a non-uniform overweight $w/w_{\mathrm{max,eff}}$.
Commonly, \(w_{\mathrm{max,eff}}\) is defined to fix the fraction of overweight events via their contribution to the integral, where e.g.\ $\epsilon_{0.001}$ denotes the unweighting efficiency for which maximally \SI{0.1}{\percent} of the integral is contributed by events with overweights.

To increase the unweighting efficiency,
automatic optimization techniques can be an effective
alternative to manually tuning \(\phi\). To implement these,
\(\phi\) is combined
with a second bijective map,
\(\psi_{\theta}: U \to U\), a parametric model with parameters \(\theta\) that can be adapted to data.
Let $q_\theta$ be the pushforward of the uniform distribution $u_d$ by $\psi_\theta$,
\begin{equation}\label{eq:push_forward_psi}
    q_\theta = (\psi_\theta)_* u_d \,.
\end{equation}
Sampling from \(q_\theta\), the unweighting efficiency becomes
\begin{equation}\label{eq:unw_eff_theta}
  \epsilon_\theta = \frac{\langle w_\theta\rangle}{w_{\text{max},\theta}}  \quad \text{with} \quad w_\theta(x) \coloneq
    \frac{p_0(x)}{q_\theta(x)}\,.
\end{equation}
In order to maximize the unweighting efficiency, \(\psi_\theta\) needs to be optimized such that \(q_\theta\) becomes
close to \(p_0\).

A simple example for \(\psi_\theta\) is \Vegas, which is a fully factorized approach. It constructs
\(\psi_\theta\) as a product deformation map based on one-dimensional piecewise linear maps. When \(p_0\) is not
factorizable, this approach is clearly limited. A more expressive alternative is given by NFs,
which use neural networks to parametrize \(\psi_\theta\). 
Through restrictions in their architecture, NFs can be designed in a way so that their Jacobian
determinant can easily be evaluated without having to invert the neural networks. Flows based on Coupling Layers, for example,
are implemented as a chain of discrete, simpler steps. 
In this work, we propose to realize \(\psi_{\theta}\) as a CNF, where the map is constructed implicitly by integrating a time-dependent vector field.

\textit{Continuous Normalizing Flows}---As discussed above, we aim to sample from a target density \(p_0: \mathbb{R}^d \to
\mathbb{R}\), but only have access to a latent density \(q_0\). We seek a map \(\psi: \mathbb{R}^d \to \mathbb{R}^d\)
that transforms \(x \sim q_0\) into \(x' \sim p_0\), or approximately so. NFs provide such a map as a
trainable diffeomorphism. Different NF methods vary in construction and training objective. One example are
discrete-time flows defined as compositions \(\psi = \psi_k \circ \cdots \circ \psi_1\), often built using Coupling
Layers and trained via KL loss~\cite{10.1214/aoms/1177729694}. We instead realize \(\psi\) as
a CNF—a continuous-time analogue—which, as shown below, trains more easily and scales better with dimensionality in our application.

A CNF is defined by a smooth, time-dependent vector field \(v_t : [0,1] \times \mathbb{R}^d \to \mathbb{R}^d\), which
determines a flow \(\psi_t:\mathbb{R}^d \to \mathbb{R}^d\) via the ODE
\begin{equation}\label{eq:ode}
    \frac{\dif \psi_t(x)}{\dif t} = v_t(\psi_t(x)) \,,
\end{equation}
with initial condition \(\psi_0(x) = x\). Sampling is done by drawing \(x_0 \sim q_0\) and integrating
\begin{equation}\label{eq:cnf_sampling}
    x_t = \psi_t(x_0) = \int_0^t v_{t'}(\psi_{t'}(x_0)) \dif t' \,.
\end{equation}
If \(v_t\) is Lipschitz in space and continuous in time, the Picard--Lindelöf theorem ensures that \(\psi_t\) is bijective.

To evaluate the density \(q_t(x_t)\),
the following continuity equation is used
\begin{equation}
    \frac{\partial}{\partial t} q_t(x) + \nabla \cdot (q_t(x) v_t(x)) = 0 \,,
\end{equation}
which leads to
\begin{equation}
    \frac{\dif}{\dif t} \log q_t(\psi_t(x_0)) + \nabla \cdot v_t(\psi_t(x_0)) = 0 \,.
\end{equation}
Thus, sampling and density evaluation reduce to solving the joint system
\begin{equation}\label{eq:joint_ode}
    \frac{\dif}{\dif t} \begin{bmatrix}
           \psi_t(x_0) \\
           \log q_t(\psi_t(x_0))
         \end{bmatrix} = \begin{bmatrix}
                           v_t(\psi_t(x_0)) \\
                           - \nabla \cdot v_t(\psi_t(x_0))
                          \end{bmatrix}
\end{equation}
with initial conditions
\begin{equation}
    \begin{bmatrix}
        \psi_t(x_0) \\
        \log q_t(x_0)
    \end{bmatrix}_{t=0} = \begin{bmatrix}
                        x_0 \\
                        \log q_0(x_0)
                    \end{bmatrix} \,.
\end{equation}

Note that we define \(\psi_t\) on \(\mathbb{R}^d\) rather than the unit cube~\(U\) to avoid boundary constraints on \(v_t\),
naturally choosing a standard normal base distribution. To map outputs to \(U\),
an element-wise sigmoid transform is applied,
adjusting densities via the Jacobian determinant.

\textit{Flow Matching}---We want to learn \(v_t\) (or \(\psi_t\)) so that the model density \(q_1\) matches the target \(p\). To this end, we parametrize the vector field as \(v_{t,\theta}\) via a neural network with trainable parameters \(\theta\). 
Assuming access to samples \(x_1 \sim p(x)\), we aim to adapt \(v_{t,\theta}\).
One option is maximum likelihood
estimation via KL minimization
as in~\cite{Chen:2018}. However, this is slow because each training step requires integrating the reverse ODE to
evaluate the density.

This motivates simulation-free alternatives~\cite{Lipman:2023,Liu:2022,Albergo:2023building,Albergo:2023stochastic},
which directly match \(v_{t,\theta}\) to a target vector field \(u_t\) generating \(p(x)\). 
There are various ways to construct an admissible \(u_t\). We briefly outline the strategy of refs.~\cite{Lipman:2023,Tong:2024} via conditional vector fields.
Let 
\begin{equation}
    u_t(x \mid x_0, x_1) = x_1 - x_0 \,,
\end{equation}
with \(x_0 \sim q_0\) and \(x_1 \sim p\) being samples from the base and target distribution, respectively. 
Clearly, a particle starting at \(x_0\) will flow to \(x_1\) along a straight line when following \(u_t(\cdot \mid x_0, x_1)\), i.e.\ 
\begin{equation}\label{eq:linear_interp}
    x_t = t x_1 + (1-t) x_0 \,.
\end{equation}
Let now
\begin{equation}\label{eq:marginal_target_vector_field}
    u_t(x) = \mathbb{E}_{\substack{(x_0, x_1) \sim \pi,\\ x_t=x}} [u_t(x \mid x_0, x_1)] \,.
\end{equation}
It can be shown~\cite{Lipman:2023,Tong:2024,Liu:2022} that the flow according to \(u_t\) transforms \(q_0\) into \(p\) as long as the marginals of the joint law \(\pi\) are \(q_0\) and \(p\). For instance, we can set \(\pi(x_0, x_1) = q_0(x_0) \cdot p(x_1)\) (independent coupling). Note that we condition the expectation to pairs \((x_0, x_1)\) where \(x_t\), eq.~\eqref{eq:linear_interp}, moves through the query point \(x\). 
It is now natural to fit \(v_{t, \theta}(x)\) to \(u_t(x)\), e.g.\ via
\begin{equation}\label{eq:FM_loss_marginal}
    \mathbb{E}_{\substack{(x_0,x_1) \sim \pi,\\ t \sim U_1}} \lVert v_{t,\theta}(x_t) - u_t(x_t) \rVert^2 \,.
\end{equation}
Note that evaluating \(u_t\) at specified points \(x\) is cumbersome due to the conditioning in
eq.~\eqref{eq:marginal_target_vector_field}. Fortunately, one can show~\cite{Lipman:2023,Tong:2024} that
eq.~\eqref{eq:FM_loss_marginal} has the same minimizers as the Flow Matching objective
\begin{equation}\label{eq:FM_loss}
    \mathcal{L}_{\text{FM}} = \mathbb{E}_{\substack{(x_0, x_1) \sim \pi,\\ t \sim U_1}} \lVert
    v_{t,\theta}(x_t) -
    u_t(x_t \mid x_0, x_1) \rVert^2 \,.
\end{equation}
It is local in space and time, so it can be evaluated fast and without solving an ODE. Also, in contrast to a KL loss,
the minimizer is unique (in terms of \(v_{t,\theta}(x)\)), which should lead to more efficient use of the model parameters.
More generally, one can add noise to the interpolation to increase
robustness~\cite{Lipman:2023,Tong:2024}. This is done by adding noise with a small
standard deviation \(\sigma_{\text{noise}}\) to the individual straight lines \(x_t\). This way, the sampled data cover
more volume. We use \(\sigma_{\text{noise}} = 10^{-4}\). 

We make two modifications to the objective. First, since the data lies in the unit hypercube \(U\),
the inverse of the sigmoid transform—namely, the logit function—is used
to map it to \(\mathbb{R}^d\). To avoid numerical issues, we combine
it with 
the affine transform \(x \mapsto x \cdot (1 - \epsilon) + \epsilon/2\) with \(\epsilon = 10^{-6}\). 
Second, we cannot sample \(x_1 \sim p\) directly; instead, we sample from our model distribution \(q_1 = (\psi_\theta)_*
q_0\).
To account for the mismatch in density, the loss, eq.~\ref{eq:FM_loss}, is multiplied by the importance weight \(w\),
resulting in a variant of Energy Conditional Flow Matching~\cite{Tong:2024}. Since high variance in the weights can
impair training in high dimensions, choosing a map \(\phi\) that minimizes variance is crucial. 
After each training iteration, the model can generate samples to allow further iterative refinement. If available, a pre-trained \Vegas\ can also provide initial samples, enhancing the overall training process.

\textit{Application}---We apply our neural network optimization method to dominant partonic channels in standard-candle processes at the \LHC: lepton and top pair production with additional jets at $\sqrt s = \SI{14}{\TeV}$. We focus on the channels $d \bar d \to e^+ e^- + ng$ and $g g \to t \bar t + ng$, which contribute the largest cross sections.
We use the \Chili\ phase space to define the mapping $\phi$
which maps the unit hypercube to the physical phase space~\cite{Bothmann:2023siu}.
The \Chili\ mapping is simple yet effective,
using a single integration channel that performs competitively
with complex multichannel-based methods for several standard LHC processes,
including the ones studied in this letter~\cite{Bothmann:2023gew}.

To evaluate the matrix elements, we use \Pepper~\cite{Bothmann:2021nch,Bothmann:2022itv,Bothmann:2023gew},
a parton-level generator optimized for complex \LHC\ processes with GPU acceleration.
It employs an internal implementation of \Chili\ and is integrated in the LHC simulation toolchain by
interfaces to the widely used event generators \Sherpa~\cite{Gleisberg:2008ta,Sherpa:2019gpd,Sherpa:2024mfk} and
\Pythia~\cite{Bierlich:2022pfr,Sjostrand:2014zea,Sjostrand:2006za,Sjostrand:2007gs}.
Matrix elements are evaluated in \Pepper\ by summing over color indices using a minimal color decomposition~\cite{Melia:2013bta,Melia:2013xok,Johansson:2015oia}, while non-vanishing helicity configurations are sampled. Normally, the helicity sampling is adapted to minimize variance, similar to adding a discrete optimization dimension to \Vegas. Under neural network optimization, we jointly optimize phase-space and helicity sampling, exploiting correlations between them, by introducing the helicities as a conditioning variable for the networks.
For this study we have added a simple and straightforward file-based interface to \Pepper\ for training and evaluating Machine Learning models,
based on the scalable LHEH5 format~\cite{Hoeche:2019rti,Bothmann:2023ozs}.

For the PDFs, we
use the \NNPDF3.0~\cite{NNPDF:2014otw} set via \LHAPDF6~\cite{Buckley:2014ana}. The renormalization and factorization scales are set to $\mu_R^2=\mu_F^2=H_T'^2/2$ for lepton pair production, and to $H_T^2/2$ for top-quark pair production~\cite{Bern:2013gka}.
The following electroweak parameters are used:
$\sin^2\theta_w=0.23155$, $\alpha=1/128.80$, $m_Z=91.1876$\,GeV, $\Gamma_Z=2.4952$\,GeV, and the top-quark mass $m_t=173.21$\,GeV. All other quarks are massless.
For lepton pairs we additionally require $66\,\mathrm{GeV} \leq m_{e^+e^-} \leq 116\,\mathrm{GeV}$.
For all massless partons we enforce the additional jet cuts $p_{T,j}>30\,\mathrm{GeV}$, $|\eta_j|<5$, and $\Delta R_{ij}>0.4$. 

All \Vegas\ grids have 100 bins per dimension, optimized over 15 steps. For the highest multiplicity, we use $\sim$$3\cdot10^8$ points, yielding $2.5\cdot10^5$ training points per non-vanishing helicity configuration in lepton-pair production and $8\cdot10^4$ for top-quark pair production.
The NFs are optimized in 8 iterations.
Initially, events are generated without remapping. 
In subsequent steps, trained flows produce inputs used by \Pepper\ to generate training data.
An embedding layer encodes helicity configurations.
Our Coupling Flows use the minimal number of layers needed to capture correlations,
with transformations based on multilayer perceptrons (MLPs),
and an early stopping criterion based on validation loss.
Our ODE Flows use MLP-based vector fields with time encoded via Fourier features~\cite{Tancik:2020}.
Both models are trained using the AdamW optimizer~\cite{adamw:2019}.
After training, the models are frozen and used to generate weighted events.

\begin{figure*}[tbp]
  \centering
  \vspace{0.15cm}
    \includegraphics[width=.48\textwidth]{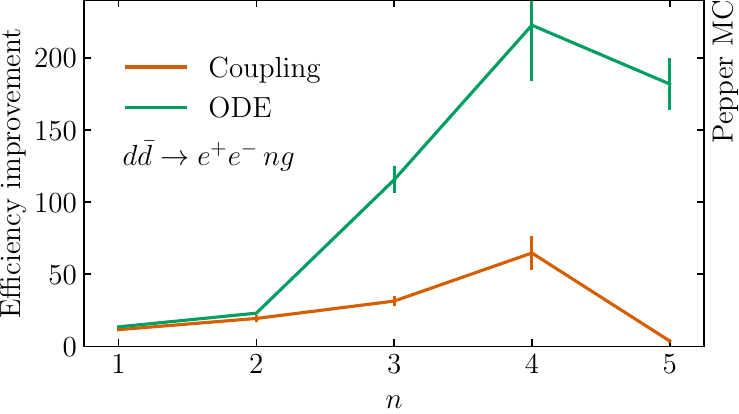}\hfill
    \includegraphics[width=.48\textwidth]{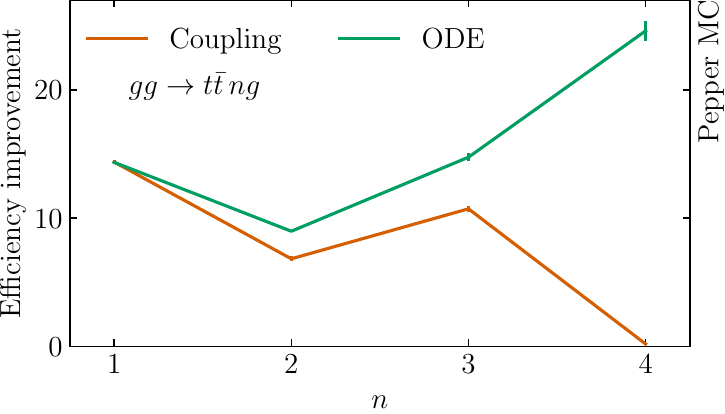}
    \caption{\label{fig:unweff}
        Relative improvements for the unweighting efficiencies $\epsilon_{0.001}$,
        both for \(e^+e^-\) + $n$ gluons (left) and for $t\bar{t}$+$n$ gluons (right),
        as a function of $n$.
        The improvements are shown for the 
        Coupling Flows (``Coupling''),
        and ODE Flows (``ODE''), compared to \Vegas. Each curve represents the mean over ten independent evaluations of the unweighting efficiency with corresponding statistical errors.}
\end{figure*}

\textit{Results}---We report relative improvements in the unweighting efficiency $\epsilon_{0.001}$, which serves as the primary performance metric of this study.  
Figure~\ref{fig:unweff} illustrates the results for the processes $d\bar d \to e^+e^-+ng$ and $gg \to t\bar t + ng$,
comparing the Coupling Flow and the ODE Flow mappings to \textsc{Vegas}.  
We observe that ODE Flows exhibit a favourable scaling of $\epsilon_{0.001}$ with increasing $n$, outperforming the other methods as the complexity grows.  
In contrast, the performance of Coupling Flows deteriorates at the highest multiplicity studied, with efficiency falling below that of the \textsc{Vegas} baseline in the $gg \to t\bar t + 4g$ case.  
For $d\bar d \to e^+e^-+5g$, the ODE Flow achieves $\epsilon_{0.001} = \SI{1.29(8)}{\percent}$, which is about 43$\times$ higher than the result for the Coupling Flow and 184$\times$ higher than the one for \Vegas.
For $\epsilon_{0.01}$, the relative gains are 13 and 153, respectively.
For $gg \to t\bar t + 4g$, the ODE Flow yields $\epsilon_{0.001} = \SI{5.76(9)}{\percent}$, outperforming the Coupling Flow by a factor of 144 and \Vegas\ by a factor of 25.
For $\epsilon_{0.01}$, the respective relative gains are 8 and 17.

Unweighting efficiency improvements of the ODE Flow over \Vegas\ are consistently greater for $d\bar d \to e^+e^-+ng$ than for $gg \to t\bar t + ng$. At $n=4$, we see a factor 198 improvement of $\epsilon_{0.001}$ for the former, versus a factor 25 for the latter. This is due to a better \Vegas\ baseline performance for top-pair production, visible also in the relative integration errors we find. This suggests structural differences in phase-space factorization---e.g.\ due to the $Z$-boson resonance and the stronger dependence on the helicity configuration in the lepton-pair production case.

Our findings demonstrate that Flow Matching outperforms alternative approaches, in particular in terms of the unweighting efficiency---the key metric for efficient event generation. Its advantage over Coupling Flows is largest at high multiplicities, where computational cost is highest, making it a highly promising candidate for current and future LHC simulation campaigns.

Although the ODE-based models incur longer inference times, their 
performance can be efficiently transferred to fast Coupling Flows using 
the recently proposed RegFlow method~\cite{Rehman:2025}, which trains the 
discrete model on pairs generated by the ODE Flow. In our benchmarks, 
RegFlow-trained Coupling Flows recover a sizable fraction of the ODE 
efficiencies while maintaining two orders of magnitude faster inference, 
yielding effective walltime gains of a factor of 12 for $d \bar d \to e^+ e^-\,5g$
and 8 for $gg \to t \bar t\,4g$ when compared to \Vegas.
The longer training is offset by the speed-up
after two million events for $d \bar d \to e^+ e^-\,5g$,
and after forty million events for $gg \to t \bar t\,4g$.

\textit{Conclusion}---We presented the first application of Flow Matching
to the problem of high-dimensional phase-space sampling in high-energy physics.
Our approach jointly remaps the input random numbers used to select continuous kinematic variables and discrete helicity configurations,
enabling more accurate and efficient sampling of complex final states.
A simple, file-based interface implemented in \Pepper, built around the \textsc{LHEH5} format, both facilitates the external optimization,
and the utilization of our results for further downstream simulation steps
using established event generation frameworks
such as \textsc{Sherpa} and \textsc{Pythia}.
Focusing on two challenging benchmark processes,
lepton-pair and top--antitop pair production with up to five and four associated jets, respectively,
we demonstrated unweighting efficiency improvements of up to 184$\times$ and 25$\times$ over \textsc{Vegas} baseline results.
We further showed that a significant fraction of the efficiency gains achieved with Flow-Matching--based ODE Flows 
can be transferred effectively to fast Coupling-Layer--based Flows using the 
RegFlow method, yielding event generation walltime reductions of around a factor of ten
at the highest jet numbers studied.
This shows that the efficiency of Coupling-Layer--based Flows
can be substantially improved by effectively using a Matching Flow objective
instead of the minimum likelihood objective otherwise used to train them in this and previous studies.
We further expect that ODE Flows themselves will become 
significantly faster as these architectures mature and more efficient 
implementations are developed.

While our study concentrated on one dominant partonic channel for each process,
full simulations will require all contributing channels.
Future work will extend the method to a single, conditional model that simultaneously
learns across multiple partonic channels and jet multiplicities, leveraging inter-channel correlations for further gains in sampling efficiency.
The sampling method improved in this way
will be made available as part of a future public \Pepper\ release,
contributing to a successful High-Luminosity LHC
and future collider physics programs.\\[0.5em]
\textit{Acknowledgments}---The authors gratefully acknowledge the computing time granted by the Resource Allocation Board and provided on the supercomputer Emmy/Grete at NHR-Nord@Göttingen as part of the NHR infrastructure. The calculations for this research were conducted with computing resources under the project nhr\_ni\_starter\_22045.
The authors also acknowledge the use of computing resources made available
by CERN to conduct some of the research reported in this work.
This material is based upon work supported by Fermi Forward Discovery Group, LLC under Contract No.\ 89243024CSC000002 with the U.S.\ Department of Energy, Office of Science, Office of High Energy Physics.
EB and MK acknowledge support by the Deutsche Forschungsgemeinschaft
(DFG, German Research Foundation) -- 510810461. TJ acknowledges financial support from the German Federal Ministry of Education and
Research (BMBF) in the ErUM-Data action plan through the KISS consortium (Verbundprojekt 05D2022). 
BS and FHS were supported by the German Research Foundation (DFG) SFB 1456, Mathematics of Experiment -- Project-ID 432680300. BS was supported by the Emmy Noether Programme of the DFG -- Project-ID 403056140.
\bibliography{literature.bib}

\begin{thebibliography}{63}%
\makeatletter
\providecommand \@ifxundefined [1]{%
 \@ifx{#1\undefined}
}%
\providecommand \@ifnum [1]{%
 \ifnum #1\expandafter \@firstoftwo
 \else \expandafter \@secondoftwo
 \fi
}%
\providecommand \@ifx [1]{%
 \ifx #1\expandafter \@firstoftwo
 \else \expandafter \@secondoftwo
 \fi
}%
\providecommand \natexlab [1]{#1}%
\providecommand \enquote  [1]{``#1''}%
\providecommand \bibnamefont  [1]{#1}%
\providecommand \bibfnamefont [1]{#1}%
\providecommand \citenamefont [1]{#1}%
\providecommand \href@noop [0]{\@secondoftwo}%
\providecommand \href [0]{\begingroup \@sanitize@url \@href}%
\providecommand \@href[1]{\@@startlink{#1}\@@href}%
\providecommand \@@href[1]{\endgroup#1\@@endlink}%
\providecommand \@sanitize@url [0]{\catcode `\\12\catcode `\$12\catcode
  `\&12\catcode `\#12\catcode `\^12\catcode `\_12\catcode `\%12\relax}%
\providecommand \@@startlink[1]{}%
\providecommand \@@endlink[0]{}%
\providecommand \url  [0]{\begingroup\@sanitize@url \@url }%
\providecommand \@url [1]{\endgroup\@href {#1}{\urlprefix }}%
\providecommand \urlprefix  [0]{URL }%
\providecommand \Eprint [0]{\href }%
\providecommand \doibase [0]{https://doi.org/}%
\providecommand \selectlanguage [0]{\@gobble}%
\providecommand \bibinfo  [0]{\@secondoftwo}%
\providecommand \bibfield  [0]{\@secondoftwo}%
\providecommand \translation [1]{[#1]}%
\providecommand \BibitemOpen [0]{}%
\providecommand \bibitemStop [0]{}%
\providecommand \bibitemNoStop [0]{.\EOS\space}%
\providecommand \EOS [0]{\spacefactor3000\relax}%
\providecommand \BibitemShut  [1]{\csname bibitem#1\endcsname}%
\let\auto@bib@innerbib\@empty
\bibitem [{\citenamefont {Aad}\ \emph {et~al.}(2008)\citenamefont {Aad} \emph
  {et~al.}}]{ATLAS:2008xda}%
  \BibitemOpen
  \bibfield  {author} {\bibinfo {author} {\bibfnamefont {G.}~\bibnamefont
  {Aad}} \emph {et~al.} (\bibinfo {collaboration} {ATLAS}),\ }\bibfield
  {title} {\bibinfo {title} {{The ATLAS Experiment at the CERN Large Hadron
  Collider}},\ }\href {https://doi.org/10.1088/1748-0221/3/08/S08003}
  {\bibfield  {journal} {\bibinfo  {journal} {JINST}\ }\textbf {\bibinfo
  {volume} {3}}\bibinfo  {number} { (08)},\ \bibinfo {pages}
  {S08003}}\BibitemShut {NoStop}%
\bibitem [{\citenamefont {Chatrchyan}\ \emph {et~al.}(2008)\citenamefont
  {Chatrchyan} \emph {et~al.}}]{CMS:2008xjf}%
  \BibitemOpen
\bibfield  {number} {  }\bibfield  {author} {\bibinfo {author} {\bibfnamefont
  {S.}~\bibnamefont {Chatrchyan}} \emph {et~al.} (\bibinfo {collaboration}
  {CMS}),\ }\bibfield  {title} {\bibinfo {title} {{The CMS Experiment at the
  CERN LHC}},\ }\href {https://doi.org/10.1088/1748-0221/3/08/S08004}
  {\bibfield  {journal} {\bibinfo  {journal} {JINST}\ }\textbf {\bibinfo
  {volume} {3}}\bibinfo  {number} { (08)},\ \bibinfo {pages}
  {S08004}}\BibitemShut {NoStop}%
\bibitem [{\citenamefont {Abada}\ \emph
  {et~al.}(2019{\natexlab{a}})\citenamefont {Abada} \emph
  {et~al.}}]{FCC:2018byv}%
  \BibitemOpen
\bibfield  {number} {  }\bibfield  {author} {\bibinfo {author} {\bibfnamefont
  {A.}~\bibnamefont {Abada}} \emph {et~al.} (\bibinfo {collaboration} {FCC}),\
  }\bibfield  {title} {\bibinfo {title} {{FCC Physics Opportunities}: {Future
  Circular Collider Conceptual Design Report Volume 1}},\ }\href
  {https://doi.org/10.1140/epjc/s10052-019-6904-3} {\bibfield  {journal}
  {\bibinfo  {journal} {Eur. Phys. J. C}\ }\textbf {\bibinfo {volume} {79}},\
  \bibinfo {pages} {474} (\bibinfo {year} {2019}{\natexlab{a}})}\BibitemShut
  {NoStop}%
\bibitem [{\citenamefont {Abada}\ \emph
  {et~al.}(2019{\natexlab{b}})\citenamefont {Abada} \emph
  {et~al.}}]{FCC:2018evy}%
  \BibitemOpen
  \bibfield  {author} {\bibinfo {author} {\bibfnamefont {A.}~\bibnamefont
  {Abada}} \emph {et~al.} (\bibinfo {collaboration} {FCC}),\ }\bibfield
  {title} {\bibinfo {title} {{FCC-ee: The Lepton Collider}: {Future Circular
  Collider Conceptual Design Report Volume 2}},\ }\href
  {https://doi.org/10.1140/epjst/e2019-900045-4} {\bibfield  {journal}
  {\bibinfo  {journal} {Eur. Phys. J. ST}\ }\textbf {\bibinfo {volume} {228}},\
  \bibinfo {pages} {261} (\bibinfo {year} {2019}{\natexlab{b}})}\BibitemShut
  {NoStop}%
\bibitem [{\citenamefont {Abada}\ \emph
  {et~al.}(2019{\natexlab{c}})\citenamefont {Abada} \emph
  {et~al.}}]{FCC:2018vvp}%
  \BibitemOpen
  \bibfield  {author} {\bibinfo {author} {\bibfnamefont {A.}~\bibnamefont
  {Abada}} \emph {et~al.} (\bibinfo {collaboration} {FCC}),\ }\bibfield
  {title} {\bibinfo {title} {{FCC-hh: The Hadron Collider}: {Future Circular
  Collider Conceptual Design Report Volume 3}},\ }\href
  {https://doi.org/10.1140/epjst/e2019-900087-0} {\bibfield  {journal}
  {\bibinfo  {journal} {Eur. Phys. J. ST}\ }\textbf {\bibinfo {volume} {228}},\
  \bibinfo {pages} {755} (\bibinfo {year} {2019}{\natexlab{c}})}\BibitemShut
  {NoStop}%
\bibitem [{\citenamefont {Buckley}\ \emph {et~al.}(2011)\citenamefont {Buckley}
  \emph {et~al.}}]{Buckley:2011ms}%
  \BibitemOpen
  \bibfield  {author} {\bibinfo {author} {\bibfnamefont {A.}~\bibnamefont
  {Buckley}} \emph {et~al.},\ }\bibfield  {title} {\bibinfo {title}
  {{General-purpose event generators for LHC physics}},\ }\href
  {https://doi.org/10.1016/j.physrep.2011.03.005} {\bibfield  {journal}
  {\bibinfo  {journal} {Phys. Rept.}\ }\textbf {\bibinfo {volume} {504}},\
  \bibinfo {pages} {145} (\bibinfo {year} {2011})},\ \Eprint
  {https://arxiv.org/abs/1101.2599} {arXiv:1101.2599 [hep-ph]} \BibitemShut
  {NoStop}%
\bibitem [{\citenamefont {{European
  Strategy~Group}}(2020)}]{EuropeanStrategyGroup:2020pow}%
  \BibitemOpen
  \bibfield  {author} {\bibinfo {author} {\bibnamefont {{European
  Strategy~Group}}},\ }\href {https://doi.org/10.17181/ESU2020} {\emph
  {\bibinfo {title} {{2020 Update of the European Strategy for Particle
  Physics}}}}\ (\bibinfo  {publisher} {CERN Council},\ \bibinfo {address}
  {Geneva},\ \bibinfo {year} {2020})\BibitemShut {NoStop}%
\bibitem [{\citenamefont {Narain}\ \emph {et~al.}(2022)\citenamefont {Narain}
  \emph {et~al.}}]{Narain:2022qud}%
  \BibitemOpen
  \bibfield  {author} {\bibinfo {author} {\bibfnamefont {M.}~\bibnamefont
  {Narain}} \emph {et~al.},\ }\href@noop {} {\bibinfo {title} {{The Future of
  US Particle Physics - The Snowmass 2021 Energy Frontier Report}}} (\bibinfo
  {year} {2022}),\ \Eprint {https://arxiv.org/abs/2211.11084} {arXiv:2211.11084
  [hep-ex]} \BibitemShut {NoStop}%
\bibitem [{\citenamefont {Albrecht}\ \emph {et~al.}(2019)\citenamefont
  {Albrecht} \emph {et~al.}}]{HEPSoftwareFoundation:2017ggl}%
  \BibitemOpen
  \bibfield  {author} {\bibinfo {author} {\bibfnamefont {J.}~\bibnamefont
  {Albrecht}} \emph {et~al.} (\bibinfo {collaboration} {HEP Software
  Foundation}),\ }\bibfield  {title} {\bibinfo {title} {{A Roadmap for HEP
  Software and Computing R\&D for the 2020s}},\ }\href
  {https://doi.org/10.1007/s41781-018-0018-8} {\bibfield  {journal} {\bibinfo
  {journal} {Comput. Softw. Big Sci.}\ }\textbf {\bibinfo {volume} {3}},\
  \bibinfo {pages} {7} (\bibinfo {year} {2019})},\ \Eprint
  {https://arxiv.org/abs/1712.06982} {arXiv:1712.06982 [physics.comp-ph]}
  \BibitemShut {NoStop}%
\bibitem [{\citenamefont {Amoroso}\ \emph {et~al.}(2021)\citenamefont {Amoroso}
  \emph {et~al.}}]{HSFPhysicsEventGeneratorWG:2020gxw}%
  \BibitemOpen
  \bibfield  {author} {\bibinfo {author} {\bibfnamefont {S.}~\bibnamefont
  {Amoroso}} \emph {et~al.} (\bibinfo {collaboration} {HSF Physics Event
  Generator WG}),\ }\bibfield  {title} {\bibinfo {title} {{Challenges in Monte
  Carlo Event Generator Software for High-Luminosity LHC}},\ }\href
  {https://doi.org/10.1007/s41781-021-00055-1} {\bibfield  {journal} {\bibinfo
  {journal} {Comput. Softw. Big Sci.}\ }\textbf {\bibinfo {volume} {5}},\
  \bibinfo {pages} {12} (\bibinfo {year} {2021})},\ \Eprint
  {https://arxiv.org/abs/2004.13687} {arXiv:2004.13687 [hep-ph]} \BibitemShut
  {NoStop}%
\bibitem [{\citenamefont {Aad}\ \emph {et~al.}(2019)\citenamefont {Aad} \emph
  {et~al.}}]{ATLAS:2019vrd}%
  \BibitemOpen
  \bibfield  {author} {\bibinfo {author} {\bibfnamefont {G.}~\bibnamefont
  {Aad}} \emph {et~al.} (\bibinfo {collaboration} {ATLAS}),\ }\bibfield
  {title} {\bibinfo {title} {{Measurement of the production cross section for a
  Higgs boson in association with a vector boson in the $H \to WW^{\ast} \to
  \ell\nu\ell\nu$ channel in $pp$ collisions at $\sqrt{s}$ = 13 TeV with the
  ATLAS detector}},\ }\href {https://doi.org/10.1016/j.physletb.2019.134949}
  {\bibfield  {journal} {\bibinfo  {journal} {Phys. Lett. B}\ }\textbf
  {\bibinfo {volume} {798}},\ \bibinfo {pages} {134949} (\bibinfo {year}
  {2019})},\ \Eprint {https://arxiv.org/abs/1903.10052} {arXiv:1903.10052
  [hep-ex]} \BibitemShut {NoStop}%
\bibitem [{\citenamefont {Aad}\ \emph {et~al.}(2021)\citenamefont {Aad} \emph
  {et~al.}}]{ATLAS:2020jwz}%
  \BibitemOpen
  \bibfield  {author} {\bibinfo {author} {\bibfnamefont {G.}~\bibnamefont
  {Aad}} \emph {et~al.} (\bibinfo {collaboration} {ATLAS}),\ }\bibfield
  {title} {\bibinfo {title} {{Measurement of the associated production of a
  Higgs boson decaying into $b$-quarks with a vector boson at high transverse
  momentum in $pp$ collisions at $\sqrt{s} = 13$ TeV with the ATLAS
  detector}},\ }\href {https://doi.org/10.1016/j.physletb.2021.136204}
  {\bibfield  {journal} {\bibinfo  {journal} {Phys. Lett. B}\ }\textbf
  {\bibinfo {volume} {816}},\ \bibinfo {pages} {136204} (\bibinfo {year}
  {2021})},\ \Eprint {https://arxiv.org/abs/2008.02508} {arXiv:2008.02508
  [hep-ex]} \BibitemShut {NoStop}%
\bibitem [{\citenamefont {Aaboud}\ \emph {et~al.}(2020)\citenamefont {Aaboud}
  \emph {et~al.}}]{ATLAS:2019zrq}%
  \BibitemOpen
  \bibfield  {author} {\bibinfo {author} {\bibfnamefont {M.}~\bibnamefont
  {Aaboud}} \emph {et~al.} (\bibinfo {collaboration} {ATLAS}),\ }\bibfield
  {title} {\bibinfo {title} {{Measurements of top-quark pair spin correlations
  in the $e\mu$ channel at $\sqrt{s} = 13$ TeV using $pp$ collisions in the
  ATLAS detector}},\ }\href {https://doi.org/10.1140/epjc/s10052-020-8181-6}
  {\bibfield  {journal} {\bibinfo  {journal} {Eur. Phys. J. C}\ }\textbf
  {\bibinfo {volume} {80}},\ \bibinfo {pages} {754} (\bibinfo {year} {2020})},\
  \Eprint {https://arxiv.org/abs/1903.07570} {arXiv:1903.07570 [hep-ex]}
  \BibitemShut {NoStop}%
\bibitem [{\citenamefont {Aad}\ \emph {et~al.}(2020)\citenamefont {Aad} \emph
  {et~al.}}]{ATLAS:2020aln}%
  \BibitemOpen
  \bibfield  {author} {\bibinfo {author} {\bibfnamefont {G.}~\bibnamefont
  {Aad}} \emph {et~al.} (\bibinfo {collaboration} {ATLAS}),\ }\bibfield
  {title} {\bibinfo {title} {{Measurement of the $t\bar{t}$ production
  cross-section in the lepton+jets channel at $\sqrt{s}=13$ TeV with the ATLAS
  experiment}},\ }\href {https://doi.org/10.1016/j.physletb.2020.135797}
  {\bibfield  {journal} {\bibinfo  {journal} {Phys. Lett. B}\ }\textbf
  {\bibinfo {volume} {810}},\ \bibinfo {pages} {135797} (\bibinfo {year}
  {2020})},\ \Eprint {https://arxiv.org/abs/2006.13076} {arXiv:2006.13076
  [hep-ex]} \BibitemShut {NoStop}%
\bibitem [{\citenamefont {Aad}\ \emph {et~al.}(2022)\citenamefont {Aad} \emph
  {et~al.}}]{ATLAS:2021yza}%
  \BibitemOpen
  \bibfield  {author} {\bibinfo {author} {\bibfnamefont {G.}~\bibnamefont
  {Aad}} \emph {et~al.} (\bibinfo {collaboration} {ATLAS}),\ }\bibfield
  {title} {\bibinfo {title} {{Modelling and computational improvements to the
  simulation of single vector-boson plus jet processes for the ATLAS
  experiment}},\ }\href {https://doi.org/10.1007/JHEP08(2022)089} {\bibfield
  {journal} {\bibinfo  {journal} {JHEP}\ }\textbf {\bibinfo {volume}
  {2022}}\bibfield  {number} {\bibinfo  {number} { (8)},\ \bibinfo {pages}
  {089}},\ }\Eprint {https://arxiv.org/abs/2112.09588} {arXiv:2112.09588
  [hep-ex]} \BibitemShut {NoStop}%
\bibitem [{\citenamefont {Bothmann}\ \emph
  {et~al.}(2022{\natexlab{a}})\citenamefont {Bothmann}, \citenamefont
  {Buckley}, \citenamefont {Christidi}, \citenamefont {G\"utschow},
  \citenamefont {H\"oche}, \citenamefont {Knobbe}, \citenamefont {Martin},\
  and\ \citenamefont {Sch\"onherr}}]{Bothmann:2022thx}%
  \BibitemOpen
  \bibfield  {author} {\bibinfo {author} {\bibfnamefont {E.}~\bibnamefont
  {Bothmann}}, \bibinfo {author} {\bibfnamefont {A.}~\bibnamefont {Buckley}},
  \bibinfo {author} {\bibfnamefont {I.~A.}\ \bibnamefont {Christidi}}, \bibinfo
  {author} {\bibfnamefont {C.}~\bibnamefont {G\"utschow}}, \bibinfo {author}
  {\bibfnamefont {S.}~\bibnamefont {H\"oche}}, \bibinfo {author} {\bibfnamefont
  {M.}~\bibnamefont {Knobbe}}, \bibinfo {author} {\bibfnamefont
  {T.}~\bibnamefont {Martin}},\ and\ \bibinfo {author} {\bibfnamefont
  {M.}~\bibnamefont {Sch\"onherr}},\ }\bibfield  {title} {\bibinfo {title}
  {{Accelerating LHC event generation with simplified pilot runs and fast
  PDFs}},\ }\href {https://doi.org/10.1140/epjc/s10052-022-11087-1} {\bibfield
  {journal} {\bibinfo  {journal} {Eur. Phys. J. C}\ }\textbf {\bibinfo {volume}
  {82}},\ \bibinfo {pages} {1128} (\bibinfo {year} {2022}{\natexlab{a}})},\
  \Eprint {https://arxiv.org/abs/2209.00843} {arXiv:2209.00843 [hep-ph]}
  \BibitemShut {NoStop}%
\bibitem [{\citenamefont {Kleiss}\ and\ \citenamefont
  {Pittau}(1994)}]{Kleiss:1994qy}%
  \BibitemOpen
  \bibfield  {author} {\bibinfo {author} {\bibfnamefont {R.}~\bibnamefont
  {Kleiss}}\ and\ \bibinfo {author} {\bibfnamefont {R.}~\bibnamefont
  {Pittau}},\ }\bibfield  {title} {\bibinfo {title} {{Weight optimization in
  multichannel Monte Carlo}},\ }\href
  {https://doi.org/10.1016/0010-4655(94)90043-4} {\bibfield  {journal}
  {\bibinfo  {journal} {Comput. Phys. Commun.}\ }\textbf {\bibinfo {volume}
  {83}},\ \bibinfo {pages} {141} (\bibinfo {year} {1994})},\ \Eprint
  {https://arxiv.org/abs/hep-ph/9405257} {arXiv:hep-ph/9405257} \BibitemShut
  {NoStop}%
\bibitem [{\citenamefont {Lepage}(1978)}]{Lepage:1977sw}%
  \BibitemOpen
  \bibfield  {author} {\bibinfo {author} {\bibfnamefont {G.~P.}\ \bibnamefont
  {Lepage}},\ }\bibfield  {title} {\bibinfo {title} {{A New Algorithm for
  Adaptive Multidimensional Integration}},\ }\href
  {https://doi.org/10.1016/0021-9991(78)90004-9} {\bibfield  {journal}
  {\bibinfo  {journal} {J. Comput. Phys.}\ }\textbf {\bibinfo {volume} {27}},\
  \bibinfo {pages} {192} (\bibinfo {year} {1978})}\BibitemShut {NoStop}%
\bibitem [{\citenamefont {Ohl}(1999)}]{Ohl:1998jn}%
  \BibitemOpen
  \bibfield  {author} {\bibinfo {author} {\bibfnamefont {T.}~\bibnamefont
  {Ohl}},\ }\bibfield  {title} {\bibinfo {title} {{Vegas revisited: Adaptive
  Monte Carlo integration beyond factorization}},\ }\href
  {https://doi.org/10.1016/S0010-4655(99)00209-X} {\bibfield  {journal}
  {\bibinfo  {journal} {Comput. Phys. Commun.}\ }\textbf {\bibinfo {volume}
  {120}},\ \bibinfo {pages} {13} (\bibinfo {year} {1999})},\ \Eprint
  {https://arxiv.org/abs/hep-ph/9806432} {arXiv:hep-ph/9806432} \BibitemShut
  {NoStop}%
\bibitem [{\citenamefont {Lepage}(2021)}]{Lepage:2020tgj}%
  \BibitemOpen
  \bibfield  {author} {\bibinfo {author} {\bibfnamefont {G.~P.}\ \bibnamefont
  {Lepage}},\ }\bibfield  {title} {\bibinfo {title} {{Adaptive multidimensional
  integration: VEGAS enhanced}},\ }\href
  {https://doi.org/10.1016/j.jcp.2021.110386} {\bibfield  {journal} {\bibinfo
  {journal} {J. Comput. Phys.}\ }\textbf {\bibinfo {volume} {439}},\ \bibinfo
  {pages} {110386} (\bibinfo {year} {2021})},\ \Eprint
  {https://arxiv.org/abs/2009.05112} {arXiv:2009.05112 [physics.comp-ph]}
  \BibitemShut {NoStop}%
\bibitem [{\citenamefont {H\"oche}\ \emph {et~al.}(2019)\citenamefont
  {H\"oche}, \citenamefont {Prestel},\ and\ \citenamefont
  {Schulz}}]{Hoche:2019flt}%
  \BibitemOpen
  \bibfield  {author} {\bibinfo {author} {\bibfnamefont {S.}~\bibnamefont
  {H\"oche}}, \bibinfo {author} {\bibfnamefont {S.}~\bibnamefont {Prestel}},\
  and\ \bibinfo {author} {\bibfnamefont {H.}~\bibnamefont {Schulz}},\
  }\bibfield  {title} {\bibinfo {title} {{Simulation of Vector Boson Plus Many
  Jet Final States at the High Luminosity LHC}},\ }\href
  {https://doi.org/10.1103/PhysRevD.100.014024} {\bibfield  {journal} {\bibinfo
   {journal} {Phys. Rev. D}\ }\textbf {\bibinfo {volume} {100}},\ \bibinfo
  {pages} {014024} (\bibinfo {year} {2019})},\ \Eprint
  {https://arxiv.org/abs/1905.05120} {arXiv:1905.05120 [hep-ph]} \BibitemShut
  {NoStop}%
\bibitem [{\citenamefont {Tabak}\ and\ \citenamefont
  {Vanden-Eijnden}(2010)}]{Tabak:2010}%
  \BibitemOpen
  \bibfield  {author} {\bibinfo {author} {\bibfnamefont {E.}~\bibnamefont
  {Tabak}}\ and\ \bibinfo {author} {\bibfnamefont {E.}~\bibnamefont
  {Vanden-Eijnden}},\ }\bibfield  {title} {\bibinfo {title} {Density estimation
  by dual ascent of the log-likelihood},\ }\href
  {https://doi.org/10.4310/CMS.2010.v8.n1.a11} {\bibfield  {journal} {\bibinfo
  {journal} {Communications in Mathematical Sciences - COMMUN MATH SCI}\
  }\textbf {\bibinfo {volume} {8}} (\bibinfo {year} {2010})}\BibitemShut
  {NoStop}%
\bibitem [{\citenamefont {Tabak}\ and\ \citenamefont
  {Turner}(2013)}]{Tabak:2013}%
  \BibitemOpen
  \bibfield  {author} {\bibinfo {author} {\bibfnamefont {E.~G.}\ \bibnamefont
  {Tabak}}\ and\ \bibinfo {author} {\bibfnamefont {C.~V.}\ \bibnamefont
  {Turner}},\ }\bibfield  {title} {\bibinfo {title} {A family of nonparametric
  density estimation algorithms},\ }\href
  {https://doi.org/https://doi.org/10.1002/cpa.21423} {\bibfield  {journal}
  {\bibinfo  {journal} {Communications on Pure and Applied Mathematics}\
  }\textbf {\bibinfo {volume} {66}},\ \bibinfo {pages} {145} (\bibinfo {year}
  {2013})}\BibitemShut {NoStop}%
\bibitem [{\citenamefont {Dinh}\ \emph {et~al.}(2015)\citenamefont {Dinh},
  \citenamefont {Krueger},\ and\ \citenamefont {Bengio}}]{Dinh:2014}%
  \BibitemOpen
  \bibfield  {author} {\bibinfo {author} {\bibfnamefont {L.}~\bibnamefont
  {Dinh}}, \bibinfo {author} {\bibfnamefont {D.}~\bibnamefont {Krueger}},\ and\
  \bibinfo {author} {\bibfnamefont {Y.}~\bibnamefont {Bengio}},\ }\bibfield
  {title} {\bibinfo {title} {{NICE:} non-linear independent components
  estimation},\ }in\ \href {http://arxiv.org/abs/1410.8516} {\emph {\bibinfo
  {booktitle} {3rd International Conference on Learning Representations,
  Workshop Track Proceedings}}}\ (\bibinfo {year} {2015})\ \Eprint
  {https://arxiv.org/abs/1410.8516} {arXiv:1410.8516 [cs.LG]} \BibitemShut
  {NoStop}%
\bibitem [{\citenamefont {Klimek}\ and\ \citenamefont
  {Perelstein}(2020)}]{Klimek:2018mza}%
  \BibitemOpen
  \bibfield  {author} {\bibinfo {author} {\bibfnamefont {M.~D.}\ \bibnamefont
  {Klimek}}\ and\ \bibinfo {author} {\bibfnamefont {M.}~\bibnamefont
  {Perelstein}},\ }\bibfield  {title} {\bibinfo {title} {{Neural Network-Based
  Approach to Phase Space Integration}},\ }\href
  {https://doi.org/10.21468/SciPostPhys.9.4.053} {\bibfield  {journal}
  {\bibinfo  {journal} {SciPost Phys.}\ }\textbf {\bibinfo {volume} {9}},\
  \bibinfo {pages} {053} (\bibinfo {year} {2020})},\ \Eprint
  {https://arxiv.org/abs/1810.11509} {arXiv:1810.11509 [hep-ph]} \BibitemShut
  {NoStop}%
\bibitem [{\citenamefont {Bothmann}\ \emph {et~al.}(2020)\citenamefont
  {Bothmann}, \citenamefont {Jan\ss{}en}, \citenamefont {Knobbe}, \citenamefont
  {Schmale},\ and\ \citenamefont {Schumann}}]{Bothmann:2020ywa}%
  \BibitemOpen
  \bibfield  {author} {\bibinfo {author} {\bibfnamefont {E.}~\bibnamefont
  {Bothmann}}, \bibinfo {author} {\bibfnamefont {T.}~\bibnamefont
  {Jan\ss{}en}}, \bibinfo {author} {\bibfnamefont {M.}~\bibnamefont {Knobbe}},
  \bibinfo {author} {\bibfnamefont {T.}~\bibnamefont {Schmale}},\ and\ \bibinfo
  {author} {\bibfnamefont {S.}~\bibnamefont {Schumann}},\ }\bibfield  {title}
  {\bibinfo {title} {{Exploring phase space with Neural Importance Sampling}},\
  }\href {https://doi.org/10.21468/SciPostPhys.8.4.069} {\bibfield  {journal}
  {\bibinfo  {journal} {SciPost Phys.}\ }\textbf {\bibinfo {volume} {8}},\
  \bibinfo {pages} {069} (\bibinfo {year} {2020})},\ \Eprint
  {https://arxiv.org/abs/2001.05478} {arXiv:2001.05478 [hep-ph]} \BibitemShut
  {NoStop}%
\bibitem [{\citenamefont {Gao}\ \emph {et~al.}(2020)\citenamefont {Gao},
  \citenamefont {H\"oche}, \citenamefont {Isaacson}, \citenamefont {Krause},\
  and\ \citenamefont {Schulz}}]{Gao:2020zvv}%
  \BibitemOpen
  \bibfield  {author} {\bibinfo {author} {\bibfnamefont {C.}~\bibnamefont
  {Gao}}, \bibinfo {author} {\bibfnamefont {S.}~\bibnamefont {H\"oche}},
  \bibinfo {author} {\bibfnamefont {J.}~\bibnamefont {Isaacson}}, \bibinfo
  {author} {\bibfnamefont {C.}~\bibnamefont {Krause}},\ and\ \bibinfo {author}
  {\bibfnamefont {H.}~\bibnamefont {Schulz}},\ }\bibfield  {title} {\bibinfo
  {title} {{Event Generation with Normalizing Flows}},\ }\href
  {https://doi.org/10.1103/PhysRevD.101.076002} {\bibfield  {journal} {\bibinfo
   {journal} {Phys. Rev. D}\ }\textbf {\bibinfo {volume} {101}},\ \bibinfo
  {pages} {076002} (\bibinfo {year} {2020})},\ \Eprint
  {https://arxiv.org/abs/2001.10028} {arXiv:2001.10028 [hep-ph]} \BibitemShut
  {NoStop}%
\bibitem [{\citenamefont {Heimel}\ \emph {et~al.}(2023)\citenamefont {Heimel},
  \citenamefont {Winterhalder}, \citenamefont {Butter}, \citenamefont
  {Isaacson}, \citenamefont {Krause}, \citenamefont {Maltoni}, \citenamefont
  {Mattelaer},\ and\ \citenamefont {Plehn}}]{Heimel:2022wyj}%
  \BibitemOpen
  \bibfield  {author} {\bibinfo {author} {\bibfnamefont {T.}~\bibnamefont
  {Heimel}}, \bibinfo {author} {\bibfnamefont {R.}~\bibnamefont
  {Winterhalder}}, \bibinfo {author} {\bibfnamefont {A.}~\bibnamefont
  {Butter}}, \bibinfo {author} {\bibfnamefont {J.}~\bibnamefont {Isaacson}},
  \bibinfo {author} {\bibfnamefont {C.}~\bibnamefont {Krause}}, \bibinfo
  {author} {\bibfnamefont {F.}~\bibnamefont {Maltoni}}, \bibinfo {author}
  {\bibfnamefont {O.}~\bibnamefont {Mattelaer}},\ and\ \bibinfo {author}
  {\bibfnamefont {T.}~\bibnamefont {Plehn}},\ }\bibfield  {title} {\bibinfo
  {title} {{MadNIS - Neural multi-channel importance sampling}},\ }\href
  {https://doi.org/10.21468/SciPostPhys.15.4.141} {\bibfield  {journal}
  {\bibinfo  {journal} {SciPost Phys.}\ }\textbf {\bibinfo {volume} {15}},\
  \bibinfo {pages} {141} (\bibinfo {year} {2023})},\ \Eprint
  {https://arxiv.org/abs/2212.06172} {arXiv:2212.06172 [hep-ph]} \BibitemShut
  {NoStop}%
\bibitem [{\citenamefont {Verheyen}(2022)}]{Verheyen:2022tov}%
  \BibitemOpen
  \bibfield  {author} {\bibinfo {author} {\bibfnamefont {R.}~\bibnamefont
  {Verheyen}},\ }\bibfield  {title} {\bibinfo {title} {{Event Generation and
  Density Estimation with Surjective Normalizing Flows}},\ }\href
  {https://doi.org/10.21468/SciPostPhys.13.3.047} {\bibfield  {journal}
  {\bibinfo  {journal} {SciPost Phys.}\ }\textbf {\bibinfo {volume} {13}},\
  \bibinfo {pages} {047} (\bibinfo {year} {2022})},\ \Eprint
  {https://arxiv.org/abs/2205.01697} {arXiv:2205.01697 [hep-ph]} \BibitemShut
  {NoStop}%
\bibitem [{\citenamefont {Heimel}\ \emph {et~al.}(2024)\citenamefont {Heimel},
  \citenamefont {Huetsch}, \citenamefont {Maltoni}, \citenamefont {Mattelaer},
  \citenamefont {Plehn},\ and\ \citenamefont {Winterhalder}}]{Heimel:2023ngj}%
  \BibitemOpen
  \bibfield  {author} {\bibinfo {author} {\bibfnamefont {T.}~\bibnamefont
  {Heimel}}, \bibinfo {author} {\bibfnamefont {N.}~\bibnamefont {Huetsch}},
  \bibinfo {author} {\bibfnamefont {F.}~\bibnamefont {Maltoni}}, \bibinfo
  {author} {\bibfnamefont {O.}~\bibnamefont {Mattelaer}}, \bibinfo {author}
  {\bibfnamefont {T.}~\bibnamefont {Plehn}},\ and\ \bibinfo {author}
  {\bibfnamefont {R.}~\bibnamefont {Winterhalder}},\ }\bibfield  {title}
  {\bibinfo {title} {{The MadNIS reloaded}},\ }\href
  {https://doi.org/10.21468/SciPostPhys.17.1.023} {\bibfield  {journal}
  {\bibinfo  {journal} {SciPost Phys.}\ }\textbf {\bibinfo {volume} {17}},\
  \bibinfo {pages} {023} (\bibinfo {year} {2024})},\ \Eprint
  {https://arxiv.org/abs/2311.01548} {arXiv:2311.01548 [hep-ph]} \BibitemShut
  {NoStop}%
\bibitem [{\citenamefont {Heimel}\ \emph {et~al.}(2025)\citenamefont {Heimel},
  \citenamefont {Mattelaer}, \citenamefont {Plehn},\ and\ \citenamefont
  {Winterhalder}}]{Heimel:2024wph}%
  \BibitemOpen
  \bibfield  {author} {\bibinfo {author} {\bibfnamefont {T.}~\bibnamefont
  {Heimel}}, \bibinfo {author} {\bibfnamefont {O.}~\bibnamefont {Mattelaer}},
  \bibinfo {author} {\bibfnamefont {T.}~\bibnamefont {Plehn}},\ and\ \bibinfo
  {author} {\bibfnamefont {R.}~\bibnamefont {Winterhalder}},\ }\bibfield
  {title} {\bibinfo {title} {{Differentiable MadNIS-Lite}},\ }\href
  {https://doi.org/10.21468/SciPostPhys.18.1.017} {\bibfield  {journal}
  {\bibinfo  {journal} {SciPost Phys.}\ }\textbf {\bibinfo {volume} {18}},\
  \bibinfo {pages} {017} (\bibinfo {year} {2025})},\ \Eprint
  {https://arxiv.org/abs/2408.01486} {arXiv:2408.01486 [hep-ph]} \BibitemShut
  {NoStop}%
\bibitem [{\citenamefont {Badger}\ \emph {et~al.}(2023)\citenamefont {Badger}
  \emph {et~al.}}]{Butter:2022rso}%
  \BibitemOpen
  \bibfield  {author} {\bibinfo {author} {\bibfnamefont {S.}~\bibnamefont
  {Badger}} \emph {et~al.},\ }\bibfield  {title} {\bibinfo {title} {{Machine
  learning and LHC event generation}},\ }\href
  {https://doi.org/10.21468/SciPostPhys.14.4.079} {\bibfield  {journal}
  {\bibinfo  {journal} {SciPost Phys.}\ }\textbf {\bibinfo {volume} {14}},\
  \bibinfo {pages} {079} (\bibinfo {year} {2023})},\ \Eprint
  {https://arxiv.org/abs/2203.07460} {arXiv:2203.07460 [hep-ph]} \BibitemShut
  {NoStop}%
\bibitem [{\citenamefont {Chen}\ \emph {et~al.}(2018)\citenamefont {Chen},
  \citenamefont {Rubanova}, \citenamefont {Bettencourt},\ and\ \citenamefont
  {Duvenaud}}]{Chen:2018}%
  \BibitemOpen
  \bibfield  {author} {\bibinfo {author} {\bibfnamefont {R.~T.~Q.}\
  \bibnamefont {Chen}}, \bibinfo {author} {\bibfnamefont {Y.}~\bibnamefont
  {Rubanova}}, \bibinfo {author} {\bibfnamefont {J.}~\bibnamefont
  {Bettencourt}},\ and\ \bibinfo {author} {\bibfnamefont {D.}~\bibnamefont
  {Duvenaud}},\ }\bibfield  {title} {\bibinfo {title} {Neural ordinary
  differential equations},\ }in\ \href@noop {} {\emph {\bibinfo {booktitle}
  {Proceedings of the 32nd International Conference on Neural Information
  Processing Systems}}}\ (\bibinfo  {publisher} {Curran Associates Inc.},\
  \bibinfo {address} {Red Hook, NY, USA},\ \bibinfo {year} {2018})\ p.\
  \bibinfo {pages} {6572–6583},\ \Eprint {https://arxiv.org/abs/1806.07366}
  {arXiv:1806.07366 [cs.LG]} \BibitemShut {NoStop}%
\bibitem [{\citenamefont {Lipman}\ \emph {et~al.}(2023)\citenamefont {Lipman},
  \citenamefont {Chen}, \citenamefont {Ben-Hamu}, \citenamefont {Nickel},\ and\
  \citenamefont {Le}}]{Lipman:2023}%
  \BibitemOpen
  \bibfield  {author} {\bibinfo {author} {\bibfnamefont {Y.}~\bibnamefont
  {Lipman}}, \bibinfo {author} {\bibfnamefont {R.~T.~Q.}\ \bibnamefont {Chen}},
  \bibinfo {author} {\bibfnamefont {H.}~\bibnamefont {Ben-Hamu}}, \bibinfo
  {author} {\bibfnamefont {M.}~\bibnamefont {Nickel}},\ and\ \bibinfo {author}
  {\bibfnamefont {M.}~\bibnamefont {Le}},\ }\bibfield  {title} {\bibinfo
  {title} {Flow matching for generative modeling},\ }in\ \href
  {https://openreview.net/forum?id=PqvMRDCJT9t} {\emph {\bibinfo {booktitle}
  {The Eleventh International Conference on Learning Representations}}}\
  (\bibinfo {year} {2023})\ \Eprint {https://arxiv.org/abs/2210.02747}
  {arXiv:2210.02747 [cs.LG]} \BibitemShut {NoStop}%
\bibitem [{\citenamefont {Albergo}\ and\ \citenamefont
  {Vanden-Eijnden}(2023)}]{Albergo:2023building}%
  \BibitemOpen
  \bibfield  {author} {\bibinfo {author} {\bibfnamefont {M.~S.}\ \bibnamefont
  {Albergo}}\ and\ \bibinfo {author} {\bibfnamefont {E.}~\bibnamefont
  {Vanden-Eijnden}},\ }\bibfield  {title} {\bibinfo {title} {Building
  normalizing flows with stochastic interpolants},\ }in\ \href
  {https://openreview.net/forum?id=li7qeBbCR1t} {\emph {\bibinfo {booktitle}
  {The Eleventh International Conference on Learning Representations}}}\
  (\bibinfo {year} {2023})\ \Eprint {https://arxiv.org/abs/2209.15571}
  {arXiv:2209.15571 [cs.LG]} \BibitemShut {NoStop}%
\bibitem [{\citenamefont {Albergo}\ \emph {et~al.}(2023)\citenamefont
  {Albergo}, \citenamefont {Boffi},\ and\ \citenamefont
  {Vanden-Eijnden}}]{Albergo:2023stochastic}%
  \BibitemOpen
  \bibfield  {author} {\bibinfo {author} {\bibfnamefont {M.~S.}\ \bibnamefont
  {Albergo}}, \bibinfo {author} {\bibfnamefont {N.~M.}\ \bibnamefont {Boffi}},\
  and\ \bibinfo {author} {\bibfnamefont {E.}~\bibnamefont {Vanden-Eijnden}},\
  }\href {https://arxiv.org/abs/2303.08797} {\bibinfo {title} {Stochastic
  interpolants: A unifying framework for flows and diffusions}} (\bibinfo
  {year} {2023}),\ \Eprint {https://arxiv.org/abs/2303.08797} {arXiv:2303.08797
  [cs.LG]} \BibitemShut {NoStop}%
\bibitem [{\citenamefont {Liu}\ \emph {et~al.}(2022)\citenamefont {Liu},
  \citenamefont {Gong},\ and\ \citenamefont {Liu}}]{Liu:2022}%
  \BibitemOpen
  \bibfield  {author} {\bibinfo {author} {\bibfnamefont {X.}~\bibnamefont
  {Liu}}, \bibinfo {author} {\bibfnamefont {C.}~\bibnamefont {Gong}},\ and\
  \bibinfo {author} {\bibfnamefont {Q.}~\bibnamefont {Liu}},\ }\href@noop {}
  {\bibinfo {title} {Flow straight and fast: Learning to generate and transfer
  data with rectified flow}} (\bibinfo {year} {2022}),\ \Eprint
  {https://arxiv.org/abs/2209.03003} {arXiv:2209.03003 [cs.LG]} \BibitemShut
  {NoStop}%
\bibitem [{\citenamefont {M\"{u}ller}\ \emph {et~al.}(2019)\citenamefont
  {M\"{u}ller}, \citenamefont {Mcwilliams}, \citenamefont {Rousselle},
  \citenamefont {Gross},\ and\ \citenamefont {Nov\'{a}k}}]{Mueller:2019}%
  \BibitemOpen
  \bibfield  {author} {\bibinfo {author} {\bibfnamefont {T.}~\bibnamefont
  {M\"{u}ller}}, \bibinfo {author} {\bibfnamefont {B.}~\bibnamefont
  {Mcwilliams}}, \bibinfo {author} {\bibfnamefont {F.}~\bibnamefont
  {Rousselle}}, \bibinfo {author} {\bibfnamefont {M.}~\bibnamefont {Gross}},\
  and\ \bibinfo {author} {\bibfnamefont {J.}~\bibnamefont {Nov\'{a}k}},\
  }\bibfield  {title} {\bibinfo {title} {Neural importance sampling},\
  }\href@noop {} {\bibfield  {journal} {\bibinfo  {journal} {ACM Trans.
  Graph.}\ }\textbf {\bibinfo {volume} {38}} (\bibinfo {year} {2019})},\
  \Eprint {https://arxiv.org/abs/1808.03856} {arXiv:1808.03856 [cs.LG]}
  \BibitemShut {NoStop}%
\bibitem [{\citenamefont {Durkan}\ \emph {et~al.}(2019)\citenamefont {Durkan},
  \citenamefont {Bekasov}, \citenamefont {Murray},\ and\ \citenamefont
  {Papamakarios}}]{Durkan:2019}%
  \BibitemOpen
  \bibfield  {author} {\bibinfo {author} {\bibfnamefont {C.}~\bibnamefont
  {Durkan}}, \bibinfo {author} {\bibfnamefont {A.}~\bibnamefont {Bekasov}},
  \bibinfo {author} {\bibfnamefont {I.}~\bibnamefont {Murray}},\ and\ \bibinfo
  {author} {\bibfnamefont {G.}~\bibnamefont {Papamakarios}},\ }\bibfield
  {title} {\bibinfo {title} {Neural spline flows},\ }in\ \href@noop {} {\emph
  {\bibinfo {booktitle} {Advances in Neural Information Processing Systems}}},\
  Vol.~\bibinfo {volume} {32},\ \bibinfo {editor} {edited by\ \bibinfo {editor}
  {\bibfnamefont {H.}~\bibnamefont {Wallach}}, \bibinfo {editor} {\bibfnamefont
  {H.}~\bibnamefont {Larochelle}}, \bibinfo {editor} {\bibfnamefont
  {A.}~\bibnamefont {Beygelzimer}}, \bibinfo {editor} {\bibfnamefont
  {F.}~\bibnamefont {d\textquotesingle Alch\'{e}-Buc}}, \bibinfo {editor}
  {\bibfnamefont {E.}~\bibnamefont {Fox}},\ and\ \bibinfo {editor}
  {\bibfnamefont {R.}~\bibnamefont {Garnett}}}\ (\bibinfo  {publisher} {Curran
  Associates, Inc.},\ \bibinfo {year} {2019})\ \Eprint
  {https://arxiv.org/abs/1906.04032} {arXiv:1906.04032 [stat.ML]} \BibitemShut
  {NoStop}%
\bibitem [{\citenamefont {Buckley}\ \emph {et~al.}(2015)\citenamefont
  {Buckley}, \citenamefont {Ferrando}, \citenamefont {Lloyd}, \citenamefont
  {Nordstr\"om}, \citenamefont {Page}, \citenamefont {R\"ufenacht},
  \citenamefont {Sch\"onherr},\ and\ \citenamefont {Watt}}]{Buckley:2014ana}%
  \BibitemOpen
  \bibfield  {author} {\bibinfo {author} {\bibfnamefont {A.}~\bibnamefont
  {Buckley}}, \bibinfo {author} {\bibfnamefont {J.}~\bibnamefont {Ferrando}},
  \bibinfo {author} {\bibfnamefont {S.}~\bibnamefont {Lloyd}}, \bibinfo
  {author} {\bibfnamefont {K.}~\bibnamefont {Nordstr\"om}}, \bibinfo {author}
  {\bibfnamefont {B.}~\bibnamefont {Page}}, \bibinfo {author} {\bibfnamefont
  {M.}~\bibnamefont {R\"ufenacht}}, \bibinfo {author} {\bibfnamefont
  {M.}~\bibnamefont {Sch\"onherr}},\ and\ \bibinfo {author} {\bibfnamefont
  {G.}~\bibnamefont {Watt}},\ }\bibfield  {title} {\bibinfo {title} {{LHAPDF6:
  parton density access in the LHC precision era}},\ }\href
  {https://doi.org/10.1140/epjc/s10052-015-3318-8} {\bibfield  {journal}
  {\bibinfo  {journal} {Eur. Phys. J. C}\ }\textbf {\bibinfo {volume} {75}},\
  \bibinfo {pages} {132} (\bibinfo {year} {2015})},\ \Eprint
  {https://arxiv.org/abs/1412.7420} {arXiv:1412.7420 [hep-ph]} \BibitemShut
  {NoStop}%
\bibitem [{\citenamefont {Kullback}\ and\ \citenamefont
  {Leibler}(1951)}]{10.1214/aoms/1177729694}%
  \BibitemOpen
  \bibfield  {author} {\bibinfo {author} {\bibfnamefont {S.}~\bibnamefont
  {Kullback}}\ and\ \bibinfo {author} {\bibfnamefont {R.~A.}\ \bibnamefont
  {Leibler}},\ }\bibfield  {title} {\bibinfo {title} {{On Information and
  Sufficiency}},\ }\href {https://doi.org/10.1214/aoms/1177729694} {\bibfield
  {journal} {\bibinfo  {journal} {{The Annals of Mathematical Statistics}}\
  }\textbf {\bibinfo {volume} {22}},\ \bibinfo {pages} {79 } (\bibinfo {year}
  {1951})}\BibitemShut {NoStop}%
\bibitem [{\citenamefont {Tong}\ \emph {et~al.}(2024)\citenamefont {Tong},
  \citenamefont {FATRAS}, \citenamefont {Malkin}, \citenamefont {Huguet},
  \citenamefont {Zhang}, \citenamefont {Rector-Brooks}, \citenamefont {Wolf},\
  and\ \citenamefont {Bengio}}]{Tong:2024}%
  \BibitemOpen
  \bibfield  {author} {\bibinfo {author} {\bibfnamefont {A.}~\bibnamefont
  {Tong}}, \bibinfo {author} {\bibfnamefont {K.}~\bibnamefont {FATRAS}},
  \bibinfo {author} {\bibfnamefont {N.}~\bibnamefont {Malkin}}, \bibinfo
  {author} {\bibfnamefont {G.}~\bibnamefont {Huguet}}, \bibinfo {author}
  {\bibfnamefont {Y.}~\bibnamefont {Zhang}}, \bibinfo {author} {\bibfnamefont
  {J.}~\bibnamefont {Rector-Brooks}}, \bibinfo {author} {\bibfnamefont
  {G.}~\bibnamefont {Wolf}},\ and\ \bibinfo {author} {\bibfnamefont
  {Y.}~\bibnamefont {Bengio}},\ }\bibfield  {title} {\bibinfo {title}
  {Improving and generalizing flow-based generative models with minibatch
  optimal transport},\ }\href {https://openreview.net/forum?id=CD9Snc73AW}
  {\bibfield  {journal} {\bibinfo  {journal} {Trans. Mach. Learn. Res.}\ }
  (\bibinfo {year} {2024})},\ \Eprint {https://arxiv.org/abs/2302.00482}
  {arXiv:2302.00482 [cs.LG]} \BibitemShut {NoStop}%
\bibitem [{\citenamefont {Bothmann}\ \emph {et~al.}(2023)\citenamefont
  {Bothmann}, \citenamefont {Childers}, \citenamefont {Giele}, \citenamefont
  {Herren}, \citenamefont {Hoeche}, \citenamefont {Isaacson}, \citenamefont
  {Knobbe},\ and\ \citenamefont {Wang}}]{Bothmann:2023siu}%
  \BibitemOpen
  \bibfield  {author} {\bibinfo {author} {\bibfnamefont {E.}~\bibnamefont
  {Bothmann}}, \bibinfo {author} {\bibfnamefont {T.}~\bibnamefont {Childers}},
  \bibinfo {author} {\bibfnamefont {W.}~\bibnamefont {Giele}}, \bibinfo
  {author} {\bibfnamefont {F.}~\bibnamefont {Herren}}, \bibinfo {author}
  {\bibfnamefont {S.}~\bibnamefont {Hoeche}}, \bibinfo {author} {\bibfnamefont
  {J.}~\bibnamefont {Isaacson}}, \bibinfo {author} {\bibfnamefont
  {M.}~\bibnamefont {Knobbe}},\ and\ \bibinfo {author} {\bibfnamefont
  {R.}~\bibnamefont {Wang}},\ }\bibfield  {title} {\bibinfo {title} {{Efficient
  phase-space generation for hadron collider event simulation}},\ }\href
  {https://doi.org/10.21468/SciPostPhys.15.4.169} {\bibfield  {journal}
  {\bibinfo  {journal} {SciPost Phys.}\ }\textbf {\bibinfo {volume} {15}},\
  \bibinfo {pages} {169} (\bibinfo {year} {2023})},\ \Eprint
  {https://arxiv.org/abs/2302.10449} {arXiv:2302.10449 [hep-ph]} \BibitemShut
  {NoStop}%
\bibitem [{\citenamefont {Bothmann}\ \emph
  {et~al.}(2024{\natexlab{a}})\citenamefont {Bothmann}, \citenamefont
  {Childers}, \citenamefont {Giele}, \citenamefont {H\"oche}, \citenamefont
  {Isaacson},\ and\ \citenamefont {Knobbe}}]{Bothmann:2023gew}%
  \BibitemOpen
  \bibfield  {author} {\bibinfo {author} {\bibfnamefont {E.}~\bibnamefont
  {Bothmann}}, \bibinfo {author} {\bibfnamefont {T.}~\bibnamefont {Childers}},
  \bibinfo {author} {\bibfnamefont {W.}~\bibnamefont {Giele}}, \bibinfo
  {author} {\bibfnamefont {S.}~\bibnamefont {H\"oche}}, \bibinfo {author}
  {\bibfnamefont {J.}~\bibnamefont {Isaacson}},\ and\ \bibinfo {author}
  {\bibfnamefont {M.}~\bibnamefont {Knobbe}},\ }\bibfield  {title} {\bibinfo
  {title} {{A portable parton-level event generator for the high-luminosity
  LHC}},\ }\href {https://doi.org/10.21468/SciPostPhys.17.3.081} {\bibfield
  {journal} {\bibinfo  {journal} {SciPost Phys.}\ }\textbf {\bibinfo {volume}
  {17}},\ \bibinfo {pages} {081} (\bibinfo {year} {2024}{\natexlab{a}})},\
  \Eprint {https://arxiv.org/abs/2311.06198} {arXiv:2311.06198 [hep-ph]}
  \BibitemShut {NoStop}%
\bibitem [{\citenamefont {Bothmann}\ \emph
  {et~al.}(2022{\natexlab{b}})\citenamefont {Bothmann}, \citenamefont {Giele},
  \citenamefont {Hoeche}, \citenamefont {Isaacson},\ and\ \citenamefont
  {Knobbe}}]{Bothmann:2021nch}%
  \BibitemOpen
  \bibfield  {author} {\bibinfo {author} {\bibfnamefont {E.}~\bibnamefont
  {Bothmann}}, \bibinfo {author} {\bibfnamefont {W.}~\bibnamefont {Giele}},
  \bibinfo {author} {\bibfnamefont {S.}~\bibnamefont {Hoeche}}, \bibinfo
  {author} {\bibfnamefont {J.}~\bibnamefont {Isaacson}},\ and\ \bibinfo
  {author} {\bibfnamefont {M.}~\bibnamefont {Knobbe}},\ }\bibfield  {title}
  {\bibinfo {title} {{Many-gluon tree amplitudes on modern GPUs: A case study
  for novel event generators}},\ }\href
  {https://doi.org/10.21468/SciPostPhysCodeb.3} {\bibfield  {journal} {\bibinfo
   {journal} {SciPost Phys. Codeb.}\ }\textbf {\bibinfo {volume} {2022}},\
  \bibinfo {pages} {3} (\bibinfo {year} {2022}{\natexlab{b}})},\ \Eprint
  {https://arxiv.org/abs/2106.06507} {arXiv:2106.06507 [hep-ph]} \BibitemShut
  {NoStop}%
\bibitem [{\citenamefont {Bothmann}\ \emph
  {et~al.}(2022{\natexlab{c}})\citenamefont {Bothmann}, \citenamefont
  {Isaacson}, \citenamefont {Knobbe}, \citenamefont {H\"oche},\ and\
  \citenamefont {Giele}}]{Bothmann:2022itv}%
  \BibitemOpen
  \bibfield  {author} {\bibinfo {author} {\bibfnamefont {E.}~\bibnamefont
  {Bothmann}}, \bibinfo {author} {\bibfnamefont {J.}~\bibnamefont {Isaacson}},
  \bibinfo {author} {\bibfnamefont {M.}~\bibnamefont {Knobbe}}, \bibinfo
  {author} {\bibfnamefont {S.}~\bibnamefont {H\"oche}},\ and\ \bibinfo {author}
  {\bibfnamefont {W.}~\bibnamefont {Giele}},\ }\bibfield  {title} {\bibinfo
  {title} {{QCD tree amplitudes on modern GPUs: A case study for novel event
  generators}},\ }\href {https://doi.org/10.22323/1.414.0222} {\bibfield
  {journal} {\bibinfo  {journal} {PoS}\ }\textbf {\bibinfo {volume}
  {ICHEP2022}},\ \bibinfo {pages} {222} (\bibinfo {year}
  {2022}{\natexlab{c}})}\BibitemShut {NoStop}%
\bibitem [{\citenamefont {Gleisberg}\ \emph {et~al.}(2009)\citenamefont
  {Gleisberg}, \citenamefont {H{\"o}che}, \citenamefont {Krauss}, \citenamefont
  {Sch{\"o}nherr}, \citenamefont {Schumann}, \citenamefont {Siegert},\ and\
  \citenamefont {Winter}}]{Gleisberg:2008ta}%
  \BibitemOpen
  \bibfield  {author} {\bibinfo {author} {\bibfnamefont {T.}~\bibnamefont
  {Gleisberg}}, \bibinfo {author} {\bibfnamefont {S.}~\bibnamefont
  {H{\"o}che}}, \bibinfo {author} {\bibfnamefont {F.}~\bibnamefont {Krauss}},
  \bibinfo {author} {\bibfnamefont {M.}~\bibnamefont {Sch{\"o}nherr}}, \bibinfo
  {author} {\bibfnamefont {S.}~\bibnamefont {Schumann}}, \bibinfo {author}
  {\bibfnamefont {F.}~\bibnamefont {Siegert}},\ and\ \bibinfo {author}
  {\bibfnamefont {J.}~\bibnamefont {Winter}},\ }\bibfield  {title} {\bibinfo
  {title} {{Event generation with SHERPA 1.1}},\ }\href
  {https://doi.org/10.1088/1126-6708/2009/02/007} {\bibfield  {journal}
  {\bibinfo  {journal} {JHEP}\ }\textbf {\bibinfo {volume} {2009}}\bibfield
  {number} {\bibinfo  {number} { (02)},\ \bibinfo {pages} {007}},\ }\Eprint
  {https://arxiv.org/abs/0811.4622} {arXiv:0811.4622 [hep-ph]} \BibitemShut
  {NoStop}%
\bibitem [{\citenamefont {Bothmann}\ \emph {et~al.}(2019)\citenamefont
  {Bothmann} \emph {et~al.}}]{Sherpa:2019gpd}%
  \BibitemOpen
  \bibfield  {author} {\bibinfo {author} {\bibfnamefont {E.}~\bibnamefont
  {Bothmann}} \emph {et~al.} (\bibinfo {collaboration} {Sherpa}),\ }\bibfield
  {title} {\bibinfo {title} {{Event Generation with Sherpa 2.2}},\ }\href
  {https://doi.org/10.21468/SciPostPhys.7.3.034} {\bibfield  {journal}
  {\bibinfo  {journal} {SciPost Phys.}\ }\textbf {\bibinfo {volume} {7}},\
  \bibinfo {pages} {034} (\bibinfo {year} {2019})},\ \Eprint
  {https://arxiv.org/abs/1905.09127} {arXiv:1905.09127 [hep-ph]} \BibitemShut
  {NoStop}%
\bibitem [{\citenamefont {Bothmann}\ \emph
  {et~al.}(2024{\natexlab{b}})\citenamefont {Bothmann} \emph
  {et~al.}}]{Sherpa:2024mfk}%
  \BibitemOpen
  \bibfield  {author} {\bibinfo {author} {\bibfnamefont {E.}~\bibnamefont
  {Bothmann}} \emph {et~al.} (\bibinfo {collaboration} {Sherpa}),\ }\bibfield
  {title} {\bibinfo {title} {{Event generation with Sherpa 3}},\ }\href
  {https://doi.org/10.1007/JHEP12(2024)156} {\bibfield  {journal} {\bibinfo
  {journal} {JHEP}\ }\textbf {\bibinfo {volume} {2024}}\bibfield  {number}
  {\bibinfo  {number} { (12)},\ \bibinfo {pages} {156}},\ }\Eprint
  {https://arxiv.org/abs/2410.22148} {arXiv:2410.22148 [hep-ph]} \BibitemShut
  {NoStop}%
\bibitem [{\citenamefont {Bierlich}\ \emph {et~al.}(2022)\citenamefont
  {Bierlich} \emph {et~al.}}]{Bierlich:2022pfr}%
  \BibitemOpen
  \bibfield  {author} {\bibinfo {author} {\bibfnamefont {C.}~\bibnamefont
  {Bierlich}} \emph {et~al.},\ }\bibfield  {title} {\bibinfo {title} {{A
  comprehensive guide to the physics and usage of PYTHIA 8.3}},\ }\href
  {https://doi.org/10.21468/SciPostPhysCodeb.8} {\bibfield  {journal} {\bibinfo
   {journal} {SciPost Phys. Codeb.}\ }\textbf {\bibinfo {volume} {2022}},\
  \bibinfo {pages} {8} (\bibinfo {year} {2022})},\ \Eprint
  {https://arxiv.org/abs/2203.11601} {arXiv:2203.11601 [hep-ph]} \BibitemShut
  {NoStop}%
\bibitem [{\citenamefont {Sj\"ostrand}\ \emph {et~al.}(2015)\citenamefont
  {Sj\"ostrand}, \citenamefont {Ask}, \citenamefont {Christiansen},
  \citenamefont {Corke}, \citenamefont {Desai}, \citenamefont {Ilten},
  \citenamefont {Mrenna}, \citenamefont {Prestel}, \citenamefont {Rasmussen},\
  and\ \citenamefont {Skands}}]{Sjostrand:2014zea}%
  \BibitemOpen
  \bibfield  {author} {\bibinfo {author} {\bibfnamefont {T.}~\bibnamefont
  {Sj\"ostrand}}, \bibinfo {author} {\bibfnamefont {S.}~\bibnamefont {Ask}},
  \bibinfo {author} {\bibfnamefont {J.~R.}\ \bibnamefont {Christiansen}},
  \bibinfo {author} {\bibfnamefont {R.}~\bibnamefont {Corke}}, \bibinfo
  {author} {\bibfnamefont {N.}~\bibnamefont {Desai}}, \bibinfo {author}
  {\bibfnamefont {P.}~\bibnamefont {Ilten}}, \bibinfo {author} {\bibfnamefont
  {S.}~\bibnamefont {Mrenna}}, \bibinfo {author} {\bibfnamefont
  {S.}~\bibnamefont {Prestel}}, \bibinfo {author} {\bibfnamefont {C.~O.}\
  \bibnamefont {Rasmussen}},\ and\ \bibinfo {author} {\bibfnamefont {P.~Z.}\
  \bibnamefont {Skands}},\ }\bibfield  {title} {\bibinfo {title} {{An
  introduction to PYTHIA 8.2}},\ }\href
  {https://doi.org/10.1016/j.cpc.2015.01.024} {\bibfield  {journal} {\bibinfo
  {journal} {Comput. Phys. Commun.}\ }\textbf {\bibinfo {volume} {191}},\
  \bibinfo {pages} {159} (\bibinfo {year} {2015})},\ \Eprint
  {https://arxiv.org/abs/1410.3012} {arXiv:1410.3012 [hep-ph]} \BibitemShut
  {NoStop}%
\bibitem [{\citenamefont {Sjostrand}\ \emph {et~al.}(2006)\citenamefont
  {Sjostrand}, \citenamefont {Mrenna},\ and\ \citenamefont
  {Skands}}]{Sjostrand:2006za}%
  \BibitemOpen
  \bibfield  {author} {\bibinfo {author} {\bibfnamefont {T.}~\bibnamefont
  {Sjostrand}}, \bibinfo {author} {\bibfnamefont {S.}~\bibnamefont {Mrenna}},\
  and\ \bibinfo {author} {\bibfnamefont {P.~Z.}\ \bibnamefont {Skands}},\
  }\bibfield  {title} {\bibinfo {title} {{PYTHIA 6.4 Physics and Manual}},\
  }\href {https://doi.org/10.1088/1126-6708/2006/05/026} {\bibfield  {journal}
  {\bibinfo  {journal} {JHEP}\ }\textbf {\bibinfo {volume} {2006}}\bibfield
  {number} {\bibinfo  {number} { (05)},\ \bibinfo {pages} {026}},\ }\Eprint
  {https://arxiv.org/abs/hep-ph/0603175} {arXiv:hep-ph/0603175} \BibitemShut
  {NoStop}%
\bibitem [{\citenamefont {Sjostrand}\ \emph {et~al.}(2008)\citenamefont
  {Sjostrand}, \citenamefont {Mrenna},\ and\ \citenamefont
  {Skands}}]{Sjostrand:2007gs}%
  \BibitemOpen
  \bibfield  {author} {\bibinfo {author} {\bibfnamefont {T.}~\bibnamefont
  {Sjostrand}}, \bibinfo {author} {\bibfnamefont {S.}~\bibnamefont {Mrenna}},\
  and\ \bibinfo {author} {\bibfnamefont {P.~Z.}\ \bibnamefont {Skands}},\
  }\bibfield  {title} {\bibinfo {title} {{A Brief Introduction to PYTHIA
  8.1}},\ }\href {https://doi.org/10.1016/j.cpc.2008.01.036} {\bibfield
  {journal} {\bibinfo  {journal} {Comput. Phys. Commun.}\ }\textbf {\bibinfo
  {volume} {178}},\ \bibinfo {pages} {852} (\bibinfo {year} {2008})},\ \Eprint
  {https://arxiv.org/abs/0710.3820} {arXiv:0710.3820 [hep-ph]} \BibitemShut
  {NoStop}%
\bibitem [{\citenamefont {Melia}(2013{\natexlab{a}})}]{Melia:2013bta}%
  \BibitemOpen
  \bibfield  {author} {\bibinfo {author} {\bibfnamefont {T.}~\bibnamefont
  {Melia}},\ }\bibfield  {title} {\bibinfo {title} {{Dyck words and multiquark
  primitive amplitudes}},\ }\href {https://doi.org/10.1103/PhysRevD.88.014020}
  {\bibfield  {journal} {\bibinfo  {journal} {Phys. Rev. D}\ }\textbf {\bibinfo
  {volume} {88}},\ \bibinfo {pages} {014020} (\bibinfo {year}
  {2013}{\natexlab{a}})},\ \Eprint {https://arxiv.org/abs/1304.7809}
  {arXiv:1304.7809 [hep-ph]} \BibitemShut {NoStop}%
\bibitem [{\citenamefont {Melia}(2013{\natexlab{b}})}]{Melia:2013xok}%
  \BibitemOpen
  \bibfield  {author} {\bibinfo {author} {\bibfnamefont {T.}~\bibnamefont
  {Melia}},\ }\bibfield  {title} {\bibinfo {title} {{Dyck words and multi-quark
  amplitudes}},\ }\href {https://doi.org/10.22323/1.197.0031} {\bibfield
  {journal} {\bibinfo  {journal} {PoS}\ }\textbf {\bibinfo {volume}
  {RADCOR2013}},\ \bibinfo {pages} {031} (\bibinfo {year}
  {2013}{\natexlab{b}})}\BibitemShut {NoStop}%
\bibitem [{\citenamefont {Johansson}\ and\ \citenamefont
  {Ochirov}(2016)}]{Johansson:2015oia}%
  \BibitemOpen
  \bibfield  {author} {\bibinfo {author} {\bibfnamefont {H.}~\bibnamefont
  {Johansson}}\ and\ \bibinfo {author} {\bibfnamefont {A.}~\bibnamefont
  {Ochirov}},\ }\bibfield  {title} {\bibinfo {title} {{Color-Kinematics Duality
  for QCD Amplitudes}},\ }\href {https://doi.org/10.1007/JHEP01(2016)170}
  {\bibfield  {journal} {\bibinfo  {journal} {JHEP}\ }\textbf {\bibinfo
  {volume} {2016}}\bibfield  {number} {\bibinfo  {number} { (01)},\ \bibinfo
  {pages} {170}},\ }\Eprint {https://arxiv.org/abs/1507.00332}
  {arXiv:1507.00332 [hep-ph]} \BibitemShut {NoStop}%
\bibitem [{\citenamefont {Höche}\ \emph {et~al.}(2019)\citenamefont {Höche},
  \citenamefont {Prestel},\ and\ \citenamefont {Schulz}}]{Hoeche:2019rti}%
  \BibitemOpen
  \bibfield  {author} {\bibinfo {author} {\bibfnamefont {S.}~\bibnamefont
  {Höche}}, \bibinfo {author} {\bibfnamefont {S.}~\bibnamefont {Prestel}},\
  and\ \bibinfo {author} {\bibfnamefont {H.}~\bibnamefont {Schulz}},\
  }\bibfield  {title} {\bibinfo {title} {{Simulation of Vector Boson Plus Many
  Jet Final States at the High Luminosity LHC}},\ }\href
  {https://doi.org/10.1103/PhysRevD.100.014024} {\bibfield  {journal} {\bibinfo
   {journal} {Phys. Rev.}\ }\textbf {\bibinfo {volume} {D100}},\ \bibinfo
  {pages} {014024} (\bibinfo {year} {2019})},\ \Eprint
  {https://arxiv.org/abs/1905.05120} {arXiv:1905.05120 [hep-ph]} \BibitemShut
  {NoStop}%
\bibitem [{\citenamefont {Bothmann}\ \emph
  {et~al.}(2024{\natexlab{c}})\citenamefont {Bothmann}, \citenamefont
  {Childers}, \citenamefont {G\"utschow}, \citenamefont {H\"oche},
  \citenamefont {Hovland}, \citenamefont {Isaacson}, \citenamefont {Knobbe},\
  and\ \citenamefont {Latham}}]{Bothmann:2023ozs}%
  \BibitemOpen
  \bibfield  {author} {\bibinfo {author} {\bibfnamefont {E.}~\bibnamefont
  {Bothmann}}, \bibinfo {author} {\bibfnamefont {T.}~\bibnamefont {Childers}},
  \bibinfo {author} {\bibfnamefont {C.}~\bibnamefont {G\"utschow}}, \bibinfo
  {author} {\bibfnamefont {S.}~\bibnamefont {H\"oche}}, \bibinfo {author}
  {\bibfnamefont {P.}~\bibnamefont {Hovland}}, \bibinfo {author} {\bibfnamefont
  {J.}~\bibnamefont {Isaacson}}, \bibinfo {author} {\bibfnamefont
  {M.}~\bibnamefont {Knobbe}},\ and\ \bibinfo {author} {\bibfnamefont
  {R.}~\bibnamefont {Latham}},\ }\bibfield  {title} {\bibinfo {title}
  {{Efficient precision simulation of processes with many-jet final states at
  the LHC}},\ }\href {https://doi.org/10.1103/PhysRevD.109.014013} {\bibfield
  {journal} {\bibinfo  {journal} {Phys. Rev. D}\ }\textbf {\bibinfo {volume}
  {109}},\ \bibinfo {pages} {014013} (\bibinfo {year} {2024}{\natexlab{c}})},\
  \Eprint {https://arxiv.org/abs/2309.13154} {arXiv:2309.13154 [hep-ph]}
  \BibitemShut {NoStop}%
\bibitem [{\citenamefont {Ball}\ \emph {et~al.}(2015)\citenamefont {Ball} \emph
  {et~al.}}]{NNPDF:2014otw}%
  \BibitemOpen
  \bibfield  {author} {\bibinfo {author} {\bibfnamefont {R.~D.}\ \bibnamefont
  {Ball}} \emph {et~al.} (\bibinfo {collaboration} {NNPDF}),\ }\bibfield
  {title} {\bibinfo {title} {{Parton distributions for the LHC Run II}},\
  }\href {https://doi.org/10.1007/JHEP04(2015)040} {\bibfield  {journal}
  {\bibinfo  {journal} {JHEP}\ }\textbf {\bibinfo {volume} {2015}}\bibfield
  {number} {\bibinfo  {number} { (04)},\ \bibinfo {pages} {040}},\ }\Eprint
  {https://arxiv.org/abs/1410.8849} {arXiv:1410.8849 [hep-ph]} \BibitemShut
  {NoStop}%
\bibitem [{\citenamefont {Bern}\ \emph {et~al.}(2013)\citenamefont {Bern},
  \citenamefont {Dixon}, \citenamefont {Febres~Cordero}, \citenamefont
  {H\"oche}, \citenamefont {Ita}, \citenamefont {Kosower}, \citenamefont
  {Ma\^\i{}tre},\ and\ \citenamefont {Ozeren}}]{Bern:2013gka}%
  \BibitemOpen
  \bibfield  {author} {\bibinfo {author} {\bibfnamefont {Z.}~\bibnamefont
  {Bern}}, \bibinfo {author} {\bibfnamefont {L.~J.}\ \bibnamefont {Dixon}},
  \bibinfo {author} {\bibfnamefont {F.}~\bibnamefont {Febres~Cordero}},
  \bibinfo {author} {\bibfnamefont {S.}~\bibnamefont {H\"oche}}, \bibinfo
  {author} {\bibfnamefont {H.}~\bibnamefont {Ita}}, \bibinfo {author}
  {\bibfnamefont {D.~A.}\ \bibnamefont {Kosower}}, \bibinfo {author}
  {\bibfnamefont {D.}~\bibnamefont {Ma\^\i{}tre}},\ and\ \bibinfo {author}
  {\bibfnamefont {K.~J.}\ \bibnamefont {Ozeren}},\ }\bibfield  {title}
  {\bibinfo {title} {{Next-to-Leading Order $W + 5$-Jet Production at the
  LHC}},\ }\href {https://doi.org/10.1103/PhysRevD.88.014025} {\bibfield
  {journal} {\bibinfo  {journal} {Phys. Rev. D}\ }\textbf {\bibinfo {volume}
  {88}},\ \bibinfo {pages} {014025} (\bibinfo {year} {2013})},\ \Eprint
  {https://arxiv.org/abs/1304.1253} {arXiv:1304.1253 [hep-ph]} \BibitemShut
  {NoStop}%
\bibitem [{\citenamefont {Tancik}\ \emph {et~al.}(2020)\citenamefont {Tancik},
  \citenamefont {Srinivasan}, \citenamefont {Mildenhall}, \citenamefont
  {Fridovich-Keil}, \citenamefont {Raghavan}, \citenamefont {Singhal},
  \citenamefont {Ramamoorthi}, \citenamefont {Barron},\ and\ \citenamefont
  {Ng}}]{Tancik:2020}%
  \BibitemOpen
  \bibfield  {author} {\bibinfo {author} {\bibfnamefont {M.}~\bibnamefont
  {Tancik}}, \bibinfo {author} {\bibfnamefont {P.~P.}\ \bibnamefont
  {Srinivasan}}, \bibinfo {author} {\bibfnamefont {B.}~\bibnamefont
  {Mildenhall}}, \bibinfo {author} {\bibfnamefont {S.}~\bibnamefont
  {Fridovich-Keil}}, \bibinfo {author} {\bibfnamefont {N.}~\bibnamefont
  {Raghavan}}, \bibinfo {author} {\bibfnamefont {U.}~\bibnamefont {Singhal}},
  \bibinfo {author} {\bibfnamefont {R.}~\bibnamefont {Ramamoorthi}}, \bibinfo
  {author} {\bibfnamefont {J.~T.}\ \bibnamefont {Barron}},\ and\ \bibinfo
  {author} {\bibfnamefont {R.}~\bibnamefont {Ng}},\ }\bibfield  {title}
  {\bibinfo {title} {Fourier features let networks learn high frequency
  functions in low dimensional domains},\ }in\ \href@noop {} {\emph {\bibinfo
  {booktitle} {Proceedings of the 34th International Conference on Neural
  Information Processing Systems}}}\ (\bibinfo  {publisher} {Curran Associates
  Inc.},\ \bibinfo {address} {Red Hook, NY, USA},\ \bibinfo {year} {2020})\
  \Eprint {https://arxiv.org/abs/2006.10739} {arXiv:2006.10739 [cs.CV]}
  \BibitemShut {NoStop}%
\bibitem [{\citenamefont {Loshchilov}\ and\ \citenamefont
  {Hutter}(2019)}]{adamw:2019}%
  \BibitemOpen
  \bibfield  {author} {\bibinfo {author} {\bibfnamefont {I.}~\bibnamefont
  {Loshchilov}}\ and\ \bibinfo {author} {\bibfnamefont {F.}~\bibnamefont
  {Hutter}},\ }\bibfield  {title} {\bibinfo {title} {Decoupled weight decay
  regularization},\ }in\ \href@noop {} {\emph {\bibinfo {booktitle} {7th
  International Conference on Learning Representations}}}\ (\bibinfo {year}
  {2019})\ \Eprint {https://arxiv.org/abs/1711.05101} {arXiv:1711.05101
  [cs.LG]} \BibitemShut {NoStop}%
\bibitem [{\citenamefont {Rehman}\ \emph {et~al.}(2025)\citenamefont {Rehman},
  \citenamefont {Davis}, \citenamefont {Lu}, \citenamefont {Tang},
  \citenamefont {Bronstein}, \citenamefont {Bengio}, \citenamefont {Tong},\
  and\ \citenamefont {Bose}}]{Rehman:2025}%
  \BibitemOpen
  \bibfield  {author} {\bibinfo {author} {\bibfnamefont {D.}~\bibnamefont
  {Rehman}}, \bibinfo {author} {\bibfnamefont {O.}~\bibnamefont {Davis}},
  \bibinfo {author} {\bibfnamefont {J.}~\bibnamefont {Lu}}, \bibinfo {author}
  {\bibfnamefont {J.}~\bibnamefont {Tang}}, \bibinfo {author} {\bibfnamefont
  {M.}~\bibnamefont {Bronstein}}, \bibinfo {author} {\bibfnamefont
  {Y.}~\bibnamefont {Bengio}}, \bibinfo {author} {\bibfnamefont
  {A.}~\bibnamefont {Tong}},\ and\ \bibinfo {author} {\bibfnamefont {A.~J.}\
  \bibnamefont {Bose}},\ }\href {https://arxiv.org/abs/2506.01158} {\bibinfo
  {title} {Efficient regression-based training of normalizing flows for
  boltzmann generators}} (\bibinfo {year} {2025}),\ \Eprint
  {https://arxiv.org/abs/2506.01158} {arXiv:2506.01158 [cs.LG]} \BibitemShut
  {NoStop}%
\end{thebibliography}%
\end{document}